\documentclass[final]{aipproc}
\usepackage{amsmath}
\layoutstyle{8x11single}

\newcommand{\tr}{\operatorname{Tr}}
\newcommand{\rmd}{{\rm d}}
\newcommand{\rme}{{\rm e}}
\newcommand{\rmi}{{\rm i}}

\begin{document}

\title{Modelling excitonic-energy transfer in light-harvesting complexes}


\author{Tobias Kramer}{
  address={Institut f\"ur Physik, Humboldt-Universit\"at zu Berlin, Germany}
  ,altaddress={Department of Physics, Harvard University, USA} 
}
\author{Christoph Kreisbeck}{
  address={Department of Chemistry and Chemical Biology, Harvard University, USA}
}

\begin{abstract}
The theoretical and experimental study of energy transfer in photosynthesis has revealed an interesting transport regime, which lies at the borderline between classical transport dynamics and quantum-mechanical interference effects. 
Dissipation is caused by the coupling of electronic degrees of freedom to vibrational modes and leads to a directional energy transfer from the antenna complex to the target reaction-center. 
The dissipative driving is robust and does not rely on fine-tuning of specific vibrational modes.
For the parameter regime encountered in the biological systems new theoretical tools are required to directly compare theoretical results with experimental spectroscopy data.
The calculations require to utilize  massively parallel graphics processor units (GPUs) for efficient and exact computations.
\end{abstract}

\maketitle

\section{Introduction}

The interplay of coherent and dissipative dynamics is the main topic of this short review with a focus on modelling energy-transfer in a biological photosynthetic complex, often referred to as light-harvesting complex (LHC).
While the study of photosynthesis has a long history, the experimental observations of long-lasting beatings in light-harvesting complexes at $T=277$~K \cite{Engel2007a,Collini2010a,Panitchayangkoon2011a} sparked a debate about the origin and prevalence of quantum-mechanical effects in larger molecular networks in thermal environments.
The key experiments are performed using 2d echo-spectroscopy, which yields a two-dimensional map in the frequency domain of the optically induced electronic excitations for various delay-times between the excitation and signal pulses.
In the 2d echo-spectra, the off-diagonal cross-peaks show long-lasting periodic oscillations which have been interpreted as signatures of electronic coherence \cite{Engel2007a}.
Such a long-lasting oscillation had not been anticipated from previous experiments which were modeled within the perturbative Redfield equations.
The Redfield equations reproduce the observed fast relaxation and thermalization of the systems within several picoseconds \cite{Brixner2005a}, but do not predict oscillations at ambient temperature.
Since 2007 many theoretical proposals have either challenged the interpretation of the experimental results by showing that within some calculations no electronic coherence prevails \cite{Olbrich2011b} or hypothesized that a few specific vibronic modes exert a dominant influence on the experimental data and the signal shows vibronic resonances rather than electronic coherence \cite{Christensson2012a,Chin2013a,Chenu2013a}, while theoretical calculations for the FMO complex with structured vibrational densities displayed only a very minor impact of added vibrational peaks \cite{Kreisbeck2012a}.
Initiated and motivated by the experiments on light-harvesting complexes, the more general notion of ``environmentally assisted transport'' was used and it was conjectured that efficient transport through a network requires both, coherence and dissipation \cite{Rebentrost2009a,Plenio2008a}.
The general interest is to understand transport through open quantum systems at high temperature and to learn lessons for artificial nanostructures which could serve to convert light to electricity and mimic parts of the photosynthetic process.
Most experimental data is available for the Fenna-Matthews-Olson (FMO) complex, whose protein structure was determined in the 70ies. The FMO complex is one of the most primitive and smallest light-harvesting complexes and contains only 7-8 bacteriocholorphylls repeatedly arranged in a trimer \cite{Fenna1975a,Tronrud2009a}. 
Optical absorption-, fluorescence-, and 2d-echo-spectra of the FMO complex are known and in combination with quantum-chemical calculations a basic understanding of the couplings between the bacteriochlorophylls exists and allows one to construct an effective Hamiltonian of the FMO complex.
The absence of any ``small parameter'' in the couplings of the electronic Hamiltonian to the the thermally excited vibrations obliterates the use of perturbation expansions and makes calculations very demanding.
In this contribution, we will review methods to compute and analyze 2d echo-spectra and discuss to what extent theory allows for quantum mechanical coherences at ambient temperatures in photosynthesis.
The last section highlights the importance of the continuous vibrational mode distribution for efficient transfer and in particular shows that a specific tuning of vibrational modes to match resonantly electronic energy-differences does not play an important role for the energy flow through the FMO complex.

\section{Setting up the model}

\begin{figure}
\includegraphics[width=0.35\textwidth]{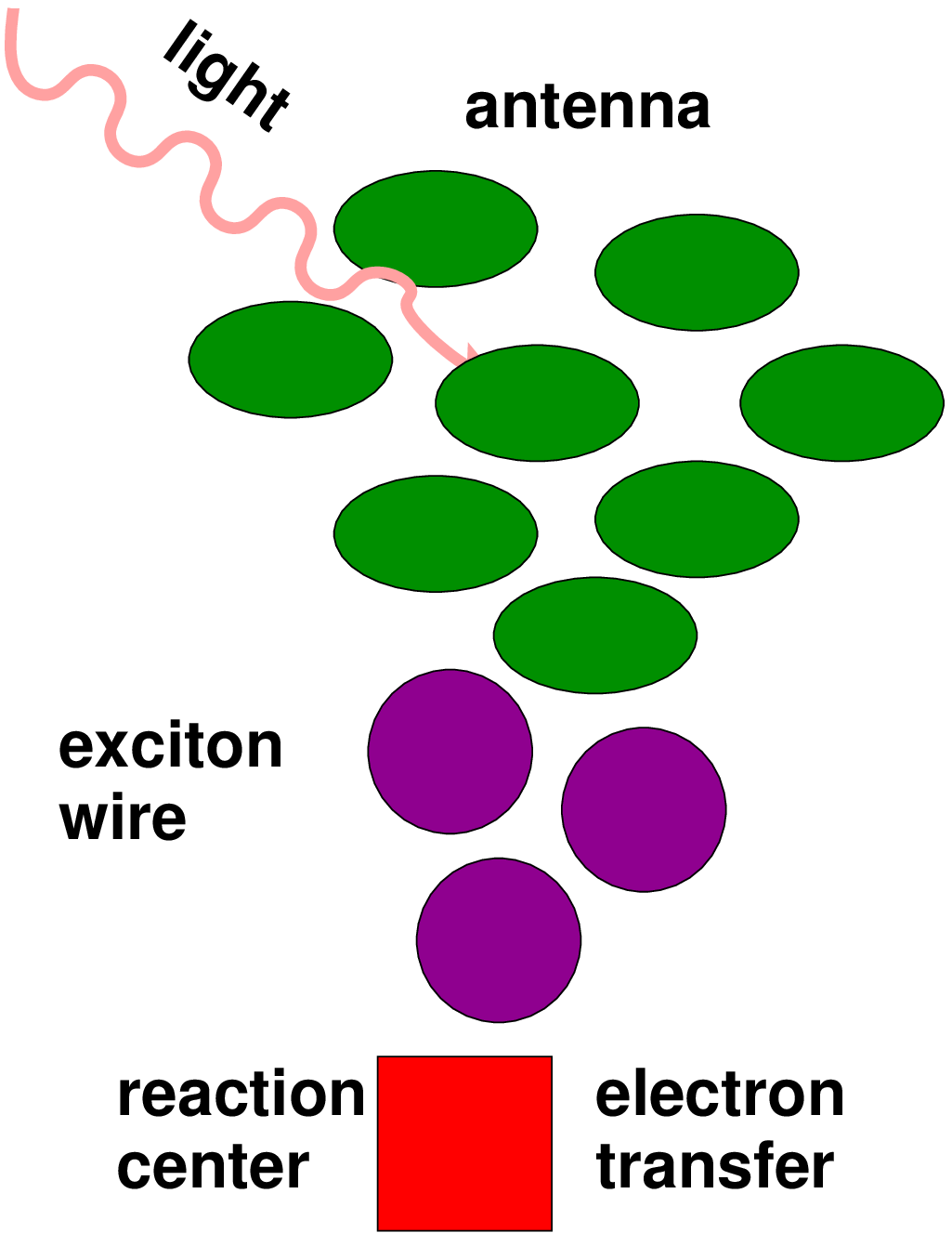} \hspace{0.2\textwidth}
\includegraphics[width=0.40\textwidth]{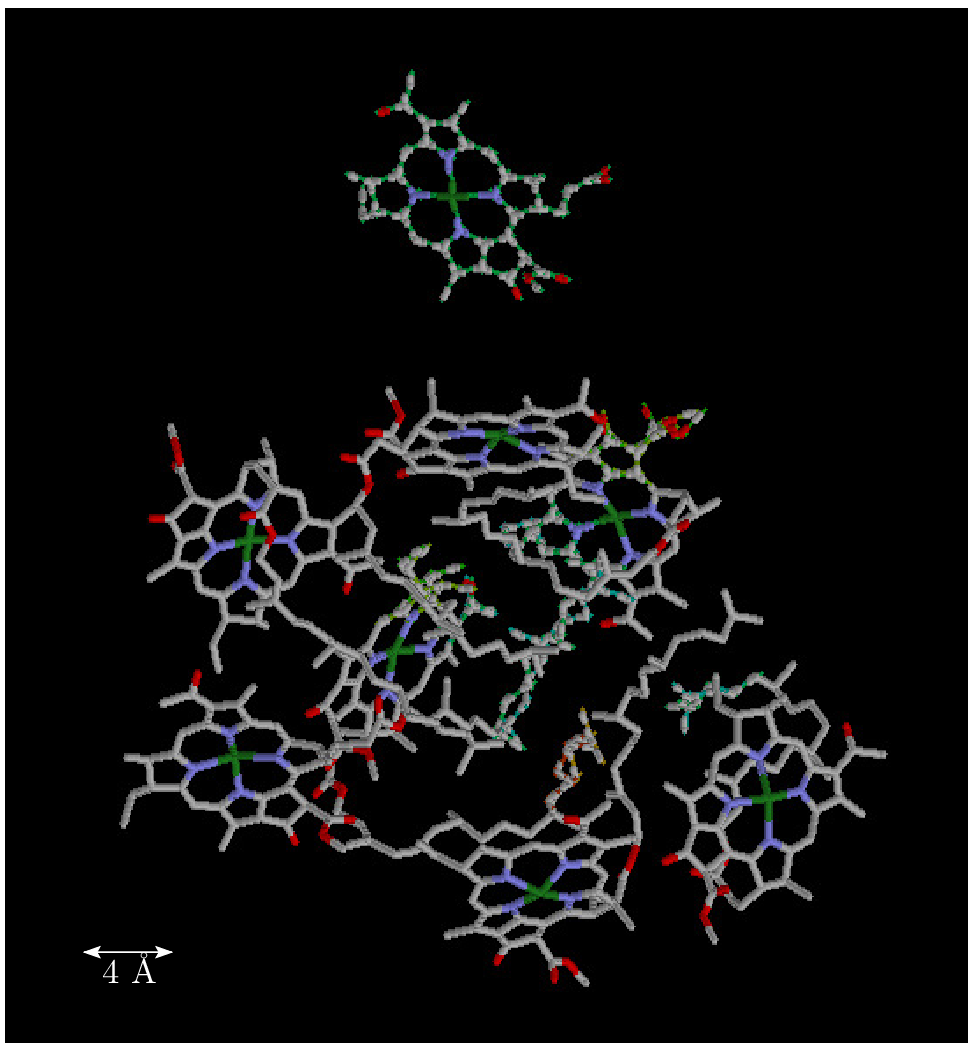}
\caption{\label{fig:greensulfur}(color online) Left panel: schematic process of light harvesting in green sulfur bacteria. Light is absorbed by the antenna and travels through the FMO network (=exciton wire) to the reaction center.
Right panel: arrangement of the eight bacteriochlorophylls in a subunit of the FMO trimer \cite{Tronrud2009a}.
}
\end{figure}
The basic process of energy transfer in photosynthesis starts with the absorption of light by the antenna part of the system. 
The light induces an electronic excitation which is transferred through the FMO complex acting as an exciton wire from the antenna to the reaction center (see Fig.~\ref{fig:greensulfur}) \cite{Li2006a}.
The collection of light with a large antenna enables the green sulfur bacteria to live under low-light conditions (for instance in the soil or in the deep sea).
In all experiments discussed here, the FMO complex has been isolated from the bacteria before spectroscopic investigation.
The FMO complex is one of the few light-harvesting complexes which has been successfully crystallized, which is a prerequisite for determining an atomistic structure, now deposited in the protein data bank (entry 3ENI \cite{Tronrud2009a}).
Three subunits constitute the FMO trimer, each subunit consists of 7 or 8 bacteriochlorophylls (Fig.~\ref{fig:greensulfur}), in the following referred to as ``sites''.
The arrangement of the bacteriochlorophylls is held in place by a protein scaffold (not shown in the figures).
The number of atoms in the FMO complex prohibits an ab-initio fully quantum mechanically simulation of the exciton transport at ambient temperatures.
As a substitute mixed classical/quantum computations \cite{Shim2012a,Aghtar2013a} and methods based on charge-density coupling \cite{Renger2012a} are used to establish an effective model of the excitonic site energies $\epsilon_m$ and the couplings between sites $J_{mn}$, leading to a seven-site model of a FMO complex subunit \cite{Renger2006a}, table~I, column~(4)
\begin{eqnarray}\label{eq:Hex}
H_{\rm ex}
&=&\sum_{m=1}^N \epsilon_m^0 |m\rangle\langle m| + \sum_{m>n} J_{mn}(|m\rangle\langle n|+|n\rangle\langle m|)\\\nonumber
&=&
\begin{pmatrix}
{12410}& -87& 5& -5& 6& -13& -9\\
-87& {12530}& 30& 8& 1& 11& 4\\
5& 30& {12210}& -53& -2& -9& 6\\
-5& 8& -53& {12320}& -70& -17& -63\\
6& 1& -2& -70& {12480}& 81& -1\\
-13& 11& -9& -17& 81& {12630}& 39\\
-9 & 4& 6& -63& -1& 39& {12440}
\end{pmatrix}\text{cm}^{-1}
\end{eqnarray}
To include dissipation and decoherence in the model, a coupling of the exciton Hamiltonian $H_{\rm ex}$ to vibrational degrees of freedom has to specified.
This is done by adding harmonic oscillators 
\begin{equation}
H_{\rm bath}=\sum_i\left(\frac{p_i^2}{2m_i}+\frac{1}{2}m\omega_i^2x_i^2\right)=\sum_i \hbar\omega_i \left(b^\dagger_i b_i+\frac{1}{2}\right)
\end{equation}
via a linear coupling of the displacement $-\sum_i f_{mi} x_i$ to each site (including the counter term $f_{mi}^2$, see \cite{Breuer2002a})
\begin{eqnarray}
H_{\rm ex-bath}
&=&\sum_{m=1}^N \frac{1}{2}m_i\omega_i^2[-2f_{mi}x_i+f_{mi}^2] |m\rangle\langle m|\\
&=&
\sum_{m=1}^N \underbrace{\sum_i\frac{1}{2}m_i\omega_i^2f_{mi}^2}_{=:\lambda_m}|m\rangle\langle m|-
\sum_{m=1}^N \sum_i  f_{mi} m_i\omega_i^2
\underbrace{\sqrt{\frac{\hbar}{2m_i\omega_i}}(b_i+b^\dagger_i)}_{x_i} |m\rangle\langle m|.
\end{eqnarray}
The counter term shifts the site energies to $\epsilon_m=\epsilon_m^0+\lambda_m$, where $\lambda_m$ denotes the reorganization energy.
If every site has the same reorganization energy, we can drop the subscript $m$ and set $\lambda_m=\lambda$.
In the exciton model of light-harvesting complexes, the bath consists of the vibrational modes of the bacteriochlorophylls.
The total Hamiltonian is written as a product of exciton-system and bath 
\begin{eqnarray}\label{eq:Htot}
H&=&
\left[ 
\sum_{m=1}^N (\epsilon^0_m+\lambda_m) |m\rangle\langle m|+ \sum_{m>n} J_{mn}(|m\rangle\langle n|+|n\rangle\langle m|) \right] \otimes 1_{\rm bath}\\\nonumber
&+&1_{\rm ex}\otimes\sum_i \hbar\omega_i (b^\dagger_i b_i+\frac{1}{2})\\\nonumber
&-& \sum_m |m\rangle\langle m| \otimes  \sum_i \hbar\omega_{mi} d_{mi} (b_i+b^\dagger_i),
\end{eqnarray}
where we introduced $d_{mi}$ in terms of $f_{mi}$.
To describe the dynamics governed by $H$ we introduce the density matrix $\rho$ of the system and bath, which evolves according to the Liouville-von Neumann equation
\begin{equation}\label{eq:lvn}
\frac{\rmd}{\rmd t}\rho(t)=-\frac{\rmi}{\hbar}\left[H,\rho(t)\right].
\end{equation}
Depending on the experimental situation, different initial conditions for the density matrix $\rho_0=\rho(t=0)$ are considered.
The discussion is simplified by moving to the interaction picture (an operator $O$ will be denoted by $\tilde{O}$ in the interaction picture) by taking the first two terms of eq.~(\ref{eq:Htot}) as reference frame. Within this frame the interaction between the exciton system and the bath denoted by $H_{\rm ex-bath}$ drives the time-evolution.
Then eq.~(\ref{eq:lvn}) reads
\begin{equation}\label{eq:lvni}
\frac{\rmd}{\rmd t}\tilde{\rho}(t)=-\frac{\rmi}{\hbar}\left[\tilde{H}_{\rm ex-bath},\tilde{\rho}(t)\right],
\end{equation}
or in integrated form
\begin{equation}\label{eq:lvnii}
\tilde{\rho}(t)=\tilde{\rho}(0)-\frac{\rmi}{\hbar}\int_0^t\rmd s \left[\tilde{H}_{\rm ex-bath},\tilde{\rho}(s)\right].
\end{equation}
Inserting the last expression into the first form and tracing over the bath yields
\begin{equation}\label{eq:lvniii}
\frac{\rmd}{\rmd t} \tr_{\rm bath} \tilde{\rho}(t)=-\frac{\rmi}{\hbar}\tr_{\rm bath}\left[\tilde{H}_{\rm ex-bath},\tilde{\rho}(0)\right]
-\frac{1}{\hbar^2}\int_0^t\rmd s \tr_{\rm bath} \left[\tilde{H}_{\rm ex-bath}(t),\left[\tilde{H}_{\rm ex-bath}(s),\tilde{\rho}(s)\right]\right].
\end{equation}
Initially we assume that the total density matrix is given as a product of the system density-matrix $\tilde{\rho}_s$ times the bath in thermal equilibrium 
\begin{equation}
\tilde{\rho}(0)=\tilde{\rho}_s(0)\otimes\tilde{\rho}_{\rm bath},\quad
\tilde{\rho}_{\rm bath}=\frac{1}{\tr_{\rm bath} \rme^{-\beta H_{\rm bath}}}\rme^{-\beta H_{\rm bath}}.
\end{equation}

\subsection{Redfield approach}
Before we discuss exact methods to solve eq.~(\ref{eq:lvni}) we outline commonly used perturbation expansions, which however fail for the parameter regime encountered in light-harvesting complexes. 
The direct comparison of exact methods with approximative ones shows in particular how electronic coherences are protected despite the presence of dissipation and decoherence.
If we make assumption (i) that the total density matrix at time $s$ can be written as a product of exciton-system and bath part \textit{at all times} $\tilde{\rho}(s)=\tilde{\rho}_s(s)\otimes\tilde{\rho}_{\rm bath}(s)$.
Then the interaction Hamiltonian in the interaction picture is best evaluated introducing the eigenvalues $E_N$ of $H_{\rm ex}+H_{\rm reorg}$
\begin{eqnarray}\label{eq:ABT}
\tilde{H}_{\rm ex-bath} &=& \sum_m \tilde{A}_m(t) \otimes  \tilde{B}_m(t)\\\label{eq:ABTii}
\tilde{A}_m(t)&=&\rme^{\rmi H_{\rm ex} t/\hbar} |m\rangle\langle m|\rme^{-\rmi H_{\rm ex} t/\hbar}=\sum_{N,M}\langle E_M|m\rangle\langle m| E_N\rangle \rme^{-\rmi \hbar (E_M-E_N)t}|E_M\rangle\langle E_N|\\
\tilde{B}_m(t)&=&-\sum_i \hbar \omega_i d_{mi} \rme^{\rmi H_{\rm bath} t/\hbar}(b_i+b^\dagger_i)\rme^{-\rmi H_{\rm bath} t/\hbar}
\end{eqnarray}
Inserting this expression into the commutator (\ref{eq:lvniii}) results in a lengthy expression which can be after some rearrangements expressed as
\begin{equation}\label{eq:lvniv}
\frac{\rmd}{\rmd t} \tr_{\rm bath} \tilde{\rho}(t)
=-\frac{1}{\hbar^2}\sum_{m,n} \int_0^t \rmd s \left(
c_{mn}(t,s)[\tilde{A}_m(t),  \tilde{A}_n(s)\tilde{\rho}_s(s)]+
c_{nm}(s,t)[\tilde{\rho}_s(s) \tilde{A}_n(s),\tilde{A}_m(s)]
\right),
\end{equation}
where the bath-correlation function appears
\begin{equation}
c_{mn}(t,s)=\left\langle  \tilde{B}_m(t) \tilde{B}_n(s) \right\rangle_{\rm thermal}.
\end{equation}
If we take the bath stationary, the bath Hamiltonian commutes with the bath density matrix and the first rhs term of eq.~(\ref{eq:lvniii}) vanishes and in addition the bath correlation function simplifies to
\begin{equation}
c_{mn}(t,s)=c_{mn}(t-s).
\end{equation}
Using the Bose-Einstein distribution $n(\omega,\beta)=1/(\rme^{\beta \hbar \omega}-1)$ and the commutator rules for the bath operators results in the explicit form of the bath correlation function 
\begin{equation}
c_{mn}(t)=\sum_{i}{\hbar^2 d_{mi}d_{ni} \omega_i^2}\left[n(\omega_i,\beta)\rme^{\rmi\omega_i t}+(n(\omega_i,\beta)+1)\rme^{-\rmi\omega_i t}\right].
\end{equation}
We neglect correlations between vibrations at different sites and thus have $c_{mn}(t)=\delta_{m,n} c_{m,m}(t)=c_m(t)$
Upon introducing the spectral density $J(\omega)$ of the bath modes
\begin{equation}
J_{m}(\omega)=\pi \sum_i \hbar^2 \omega_i^2 d_{mi}^2 \delta(\omega-\omega_i),
\end{equation}
the correlation function can be expressed as 
\begin{eqnarray}
c_m(t)=\frac{1}{\pi} \int_0^\infty\;J_m(\omega) \left[n(\omega,\beta)\rme^{\rmi\omega t}+(n(\omega,\beta)+1)\rme^{-\rmi\omega t}\right].
\end{eqnarray}
The last expression generalizes the spectral density to a continuum distribution of vibrational bath modes, which arises in larger environments.
It is also convenient to introduce the real and imaginary parts of the correlation function defined by
\begin{eqnarray}
   S_m(t) &=& {\rm Re}[c_m(t)] = \frac{\hbar}{\pi} \int_0^\infty\rmd\omega\;J_m(\omega) \cos(\omega t) \coth\left(\frac{\beta\hbar\omega}{2}\right), \\
\chi_m(t) &=& -\frac{2}{\hbar} {\rm Im}[c_m(t)] = \frac{2}{\pi} \int_0^\infty\rmd\omega\;J_m(\omega) \sin(\omega t).
\end{eqnarray}
Under the assumption (ii) of a fast decay of the bath correlation time a time local form arises if we approximate $\tilde{\rho}(t-s')\approx\tilde{\rho}(t)$. 
This results in the time-evolution equation within the Born-Redfield approximation 
\begin{equation}\label{eq:lvnv}
\frac{\rmd}{\rmd t} \tr_{\rm bath} \tilde{\rho}(t)=
-\frac{1}{\hbar^2}\int_0^t\rmd s \tr_{\rm bath} \left[\tilde{H}_{\rm ex-bath}(t),\left[\tilde{H}_{\rm ex-bath}(s),\tilde{\rho}_s(t)\otimes\rho_{\rm bath}\right]\right].
\end{equation}
Finally we substitute $\tau=t-s$ and extend the range of integration to infinity 
\begin{eqnarray}\label{eq:lvnvi}
\frac{\rmd}{\rmd t} \tr_{\rm bath} \tilde{\rho}(t)
&=&
-\frac{1}{\hbar^2}\int_0^\infty\rmd \tau \tr_{\rm bath} \left[\tilde{H}_{\rm ex-bath}(t),\left[\tilde{H}_{\rm ex-bath}(t-\tau),\tilde{\rho}_s(t)\otimes\rho_{\rm bath}\right]\right]\\
&=& 
-\frac{1}{\hbar^2}\sum_{m,n} \int_0^\infty \rmd \tau \left(
c_{mn}(\tau)[\tilde{A}_m(t),  \tilde{A}_n(t-\tau)\tilde{\rho}_s(t)]+
c_{nm}(-\tau)[\tilde{\rho}_s(t) \tilde{A}_n(t-\tau),\tilde{A}_m(t)]
\right)
\end{eqnarray}
Using the known time-evolution (\ref{eq:ABTii}) this expression can be integrated and leads to the Redfield equations.
If one in addition applies the rotating wave approximation to the time-dependent terms one arrives at the secular Redfield equations.

\begin{figure}
\includegraphics[width=0.7\textwidth]{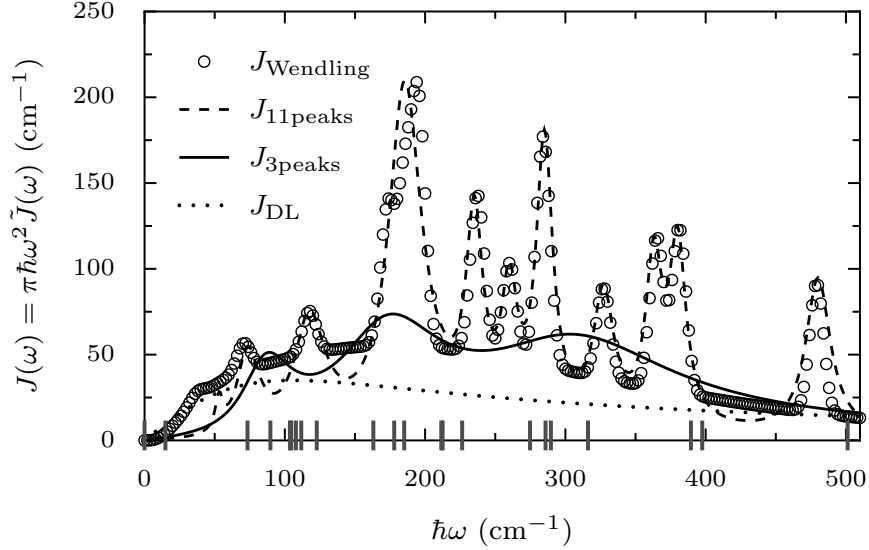}
\caption{\label{fig:specdens}Spectral density models of the FMO complex. The circles denote the experimentally extracted $J_{\rm Wendling}$ \cite{Wendling2002a}, $J_{DL}$ is a single-peak Drude-Lorentz spectral density (\ref{eq:JDL}) with $\lambda=35$~cm$^{-1}$, $\nu^{-1}=50$~fs.
The marks on the horizontal frequency axis indicate the energy difference between two exciton eigenstates of the FMO complex, where the spectral density is evaluated within the Redfield approach.
Parameters are given in Ref.~\cite{Kreisbeck2012a}.
}
\end{figure}

\subsection{Exact propagation with the hierarchical equations of motion}

The hierarchical equations of motion (HEOM) method was developed by Kubo and Tanimura to solve the time-evolution of the density matrix in an exact fashion without making the Born-Markov approximation or assuming a factorization of the total density matrix into separable system and bath density-matrices at all times. 
To obtain the hierarchy consisting of a  system of coupled differential equations requires to take a Drude-Lorentz shape of the spectral density for the bath oscillators (an restriction which we will lift later).
The Drude-Lorentz shape of spectral density is given by
\begin{equation}\label{eq:JDL}
J_{DL}(\omega)=\frac{2\lambda\omega\nu}{\omega^2+\nu^2}
\end{equation}
an explicit evaluation of the bath correlation functions yields
\begin{eqnarray}
S(t)&=&\lambda\nu\cot\left(\frac{\beta\hbar\nu}{2}\right)\rme^{-\nu t}+\frac{2\lambda}{\beta\hbar}\sum_{k=1}^\infty\frac{2\nu\mu_k\rme^{-\mu_k t}}{\mu_k^2-\nu^2},\quad \mu_k=2\pi k/\beta \\
\chi(t)&=&\frac{2\lambda\nu}{\hbar}\rme^{-\nu t}
\end{eqnarray}
The first expression is simplified for higher temperatures when the Matsubara frequencies $\mu_k$ are larger than the decay constant $\nu$, resulting in the requirement $\hbar\nu\beta\ll1$.
Replacing the exponential decay of the Matsubara frequency with a $\delta(t)$ function and keeping only the largest contribution yields the high-temperature approximation of the correlation function \cite{Ishizaki2009a}
\begin{equation}
S(t)\approx\frac{2\lambda}{\beta\hbar}\left[\left(1-\frac{2\nu^2}{\mu_1^2-\nu^2}\right)\rme^{-\nu t}+\frac{2\nu}{\mu_1^2-\nu^2}\delta(t) \right].
\end{equation} 
This form shows that the correlation functions have an exponential time-dependence, which Tanimura and Kubo exploited for a systematic expansion of the time-evolution equations in terms of hierarchically coupled equations. For a derivation we refer to \cite{Tanimura1989a,Ishizaki2009a}. 
Here we quote the result for a generalized spectral density, which is given in terms of a superposition of shifted Drude-Lorentzian peaks 
\cite{Kreisbeck2012a,Kreisbeck2013a}
\begin{equation} \label{eq:SDL}
 J(\omega)=\sum_{k=1}^{M}\left[\frac{\nu_k\lambda_k\omega}{\nu_k^2+(\omega+\Omega_k)^2}+\frac{\nu_k\lambda_k\omega}{\nu_k^2+(\omega-\Omega_k)^2}\right].
\end{equation}
Another possible decomposition is discussed in Ref.~\cite{Tanimura1994a}.
The time evolution of the reduced density operator reads
\begin{equation}\label{eq:HEOM}
 \frac{\rmd}{\rmd t}\rho(t)=
-\frac{\rmi}{\hbar}\mathcal{L}_{\rm ex}\rho(t)
+\sum_{m,k=1,s=\pm 1}^{N,M}\frac{\rmi}{\hbar} V_m^{\times} \sigma^{\vec{n}_0}(t)
\end{equation}
with
\begin{eqnarray}\label{eq:9}
\frac{\rmd}{\rmd t}\sigma^{\vec{n}}(t)&=&\mathcal{L}_{\rm ex}
\sigma^{\vec{n}}(t)\nonumber\\
&&-\sum_{m,k=1,s=\pm 1}^{N,M} n_{m,k,s}(\nu_k+s\,\rmi\Omega_k)
\big)\sigma^{\vec{n}}(t)\nonumber\\
&&+\sum_{m,k=1,s=\pm 1}^{N,M}\hspace{-0.1cm}
\big(\frac{\rmi}{\hbar}V_m^{\times}
\sigma^{\vec{n}_{m,k,s}^+}(t)+\theta_{m,k,s}\sigma^{\vec{n}_{m,k,s}^-}(t)\big). 
\end{eqnarray}
Here, we define operators
\begin{equation}
\mathcal{L}_{\rm ex}\rho=-\frac{\rmi}{\hbar}[\mathcal{H}_{\rm ex},\rho]-\sum_{m=1}^{N}\sum_{k=1}^{M}\sum_{ s=\pm 1}
\frac{2}{\beta\hbar^2}\frac{\lambda_k\nu_k}{{(\gamma_1+\rmi s\Omega_k)}^2-\nu_k^2}V_{m}^{\times}V_{m}^{\times}\rho
\end{equation}
and
\begin{equation}\label{eq:10}
\theta_{m,k,s}=\frac{\rmi}{2}\Big(\frac{2\lambda_k}{k_B T\hbar}V_{m}^{\times}
-\rmi\lambda_k(\nu_k+s\,\rmi\Omega_k) V_{m}^{\circ}-\frac{2\lambda_k}{\beta\hbar^2}\frac{{(\nu_k+\rmi s\Omega_k)}^2}{\gamma_1^2-{(\nu_k+\rmi s\Omega_k)}^2}V_{m}^{\times}\Big)
\end{equation}
with $V_{m}^{\times}\sigma=[a_m^\dag a_m, \sigma]$ and $V_{m}^{\circ}\sigma=[a_m^\dag a_m, \sigma]_+$.
The index $\vec{n}$ of the auxiliary matrix $\sigma^{\vec{n}}$ is defined
as tuple $\vec{n}=(n_{1,1,s}, ..., n_{1,M,s},..., n_{N,1,s}, ... , n_{N,M,s})$ where 
$n_{m,k,s}\in\textbf{N}$. The label $m=1,...,N$ covers the number of sites, the label
$k=1,...,M$ covers the number of shifted Drude-Lorentz peaks
and the label $s=\pm$ takes into account the sign of the frequency shift $\pm\Omega_k$. 
We further define $\vec{n}_{m,k,s}^{\pm}=(n_{1,k,s}, ..., n_{m,k,s}\pm1, ..., n_{N,M,s})$.
The hierarchy converges for a sufficiently large truncation level $N_{\rm max}=\sum_{m,k,s}n_{m,k,s}$. 
Convergence of the hierarchy must be verified by comparison with higher truncation levels.

\subsection{Comparison with other computational methods}

The parameters found in light-harvesting complexes mandate the use of exact propagation methods if one wants to study the role of quantum-mechanical coherences. 
The eigenenergy differences found in the FMO complex are about $70-500$~cm$^{-1}$, the thermal energy is around $300$~K$\times k_B\approx 200$~cm$^{-1}$, and the reorganization energy $\lambda\approx 40$~cm$^{-1}$. 
A perturbation expansion for small $\lambda$ as done within the Redfield approach fails for these parameters, and methods developed for zero temperature ($0$~K) such as DMRG \cite{Chin2013a} are not computationally suitable for the biologically relevant finite temperature.
Several exact methods are in principle available and have been tested on small systems.
The main obstacle for their application to the interpretation of experimental data and the realistic modelling of excitonic energy transfer in light-harvesting systems is the required scalability to the relevant multi-site systems.

\begin{figure}[t]
\includegraphics[width=0.5\textwidth]{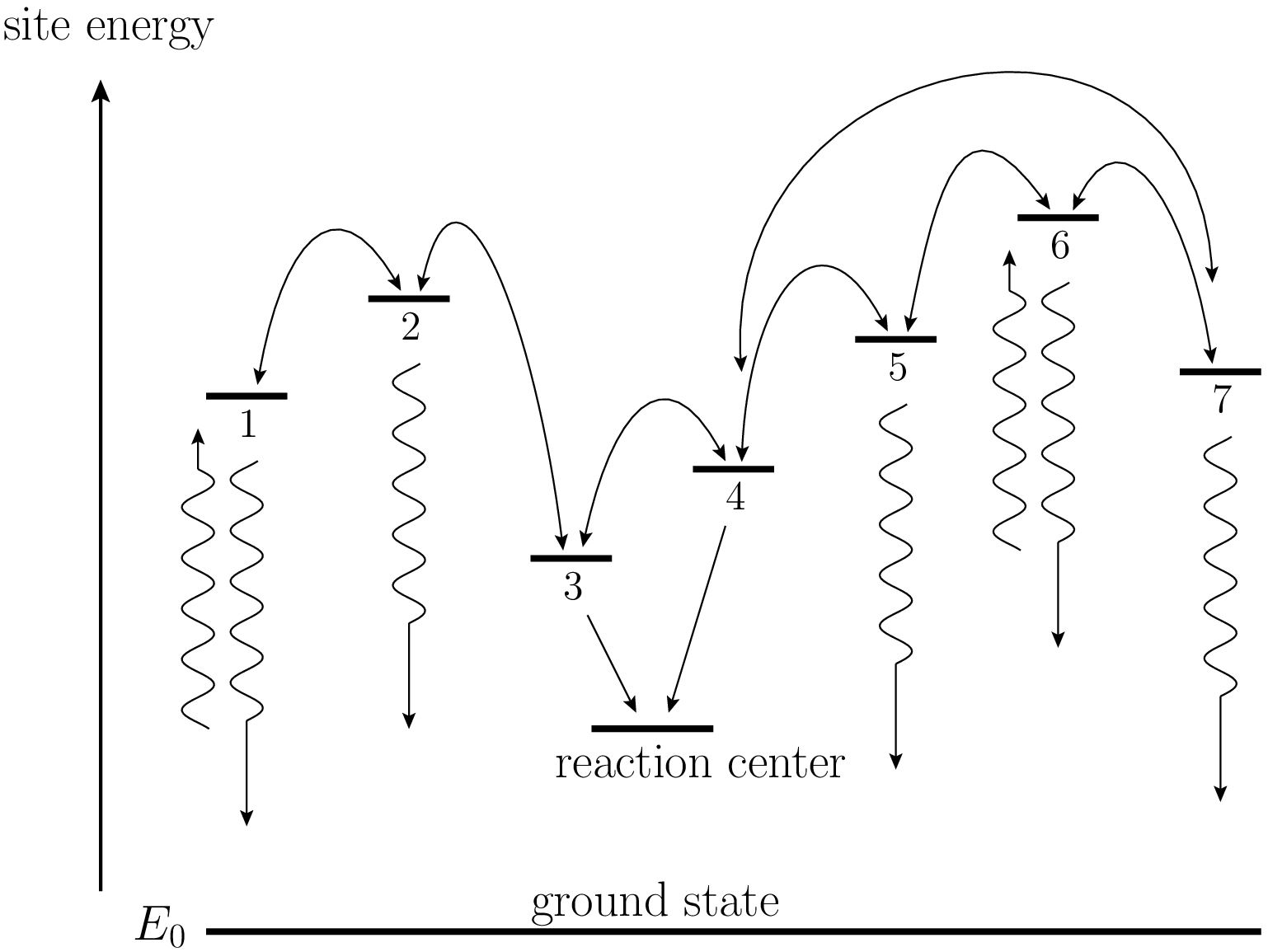}
~~~ ~~~
\includegraphics[width=0.45\textwidth]{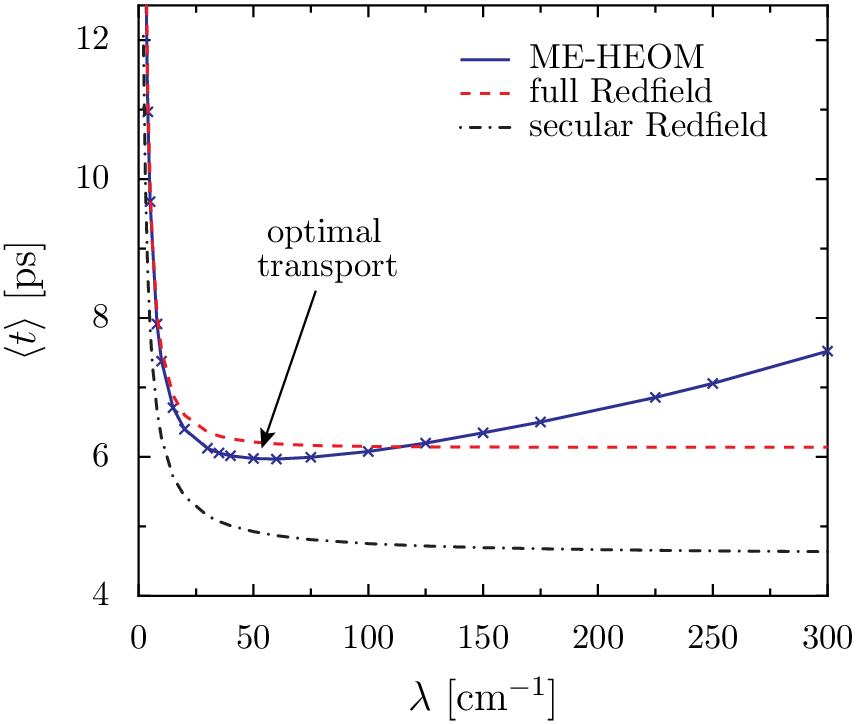}
\caption{\label{fig:fmolevels}Left panel: site energies and couplings of the seven site FMO complex model. 
Right panel: transfer time according to eq.~(\ref{eq:transfertime}) from initial site 1 to the reaction center as a function of the coupling to the spectral density parametrized by the reorganization energy $\lambda$, other parameters $\nu^{-1}=166$~fs, $T=300$~K.
The ME-HEOM result is an exact solution of the density-matrix propagation, while the Redfield results are obtained in a perturbation expansion.
}
\end{figure}

\subsubsection{Hierarchical equations of motion (HEOM)}
The hierarchical equations of motions provide important reference results for the dynamics in a regime of larger couplings \cite{Ishizaki2009a}.
The at first glance tremendous overhead imposed by the need to propagate a complete hierarchy of coupled auxiliary density matrices has been turned in a computational asset by working with massively parallel computer processors found in graphics processing units (GPU) \cite{Kreisbeck2011a,Kreisbeck2013a}.
GPUs are multicore processors and have been applied to solve the time-dependent Schr\"odinger equation \cite{Broin2012a} and to obtain the self-consistent electron flow through Hall-devices \cite{Kramer2009c,Kramer2010d} of several thousand interacting particles.
For the case of HEOM the GPU algorithm (GPU-HEOM) allows one to assign one thread out of hundreds of threads launched simultaneously to each hierarchy element and to move the complete hierarchy forward in lock-step.
Compared to deployments of the HEOM algorithm on large computer clusters \cite{Chen2011a,Struempfer2012a}, the GPU-HEOM method works on a single device and thus eliminates any network-traffic and synchronization overhead.
The high performance of the GPU-HEOM code makes it the method of choice for computing 2d echo-spectra of the FMO complex \cite{Hein2012a,Kreisbeck2012a}.
Further advantages of HEOM include:
\begin{itemize}
\item efficient implementation for parallel computers available (GPU-HEOM \cite{Kreisbeck2011a,Kreisbeck2013a}) and conventional CPU clusters \cite{Struempfer2012a}
\item reliable convergence control by systematically increasing the truncation level \cite{Tanimura1994a}
\item extendable to time-dependent system Hamiltonian \cite{Kato2013a} 
\item only non-perturbative method which has been used to compute 2d spectra of the FMO complex \cite{Chen2011a,Hein2012a,Kreisbeck2012a}
\item structured spectral densities via decomposition into shifted Drude-Lorentzian peaks \cite{Tanimura1994a,Kreisbeck2012a,Kreisbeck2013a}
\item biologically relevant temperature range requires low truncation level, 
only cryogenic temperatures (below $50-100$~K) lead to large computer memory consumption
\item ready-to-run GPU-HEOM tool online available \url{https://nanohub.org/tools/gpuheompop}
\end{itemize}
The GPU-HEOM method has been directly compared with other approaches, such as the Trotter expansion of the path-integral and a sampling of  trajectories with Monte-Carlo methods \cite{Olsina2013a}.

\section{Optimal transport}

The seven site FMO complex is an important test case for studying energy-transfer through a network under the influence of a dissipative environment. 
It has been suggested that the efficiency of the transport from the antenna to the reaction center is enhanced by the thermal environment \cite{Plenio2008a,Rebentrost2009a}.
We proceed to investigate the interplay between coherent transport (indicated by a superposition of the excitonic state over several sites and causing Rabi-type oscillations) with a relaxation process toward the thermal equilibrium.
The formalism is based on the density-matrix description reviewed in the first section.
The layout of the transport network, which is encoded in the Hamiltonian (\ref{eq:Hex}), and the strength of the coupling of the exciton system to the bath determine the transport efficiency.
A schematic representation of the FMO network is given in Fig.~\ref{fig:fmolevels}. The solid lines denote coupling elements between the seven sites which are arranged according to the site energies give in eq.~(\ref{eq:Hex}).
The initial excitation enters at site one or site six and travels through the FMO complex.
On its way to the reaction center (RC), it either vanishes (loss channel, indicated by wiggly downward arrows) or reaches the excitation-trapping reaction-center via sites three and four (straight arrows). 
The energy-levels are predominantly arranged in a funnel structure, with the lowest-lying states spatially located close to the reaction center. 
\begin{figure}[t]
\includegraphics[width=0.6\textwidth]{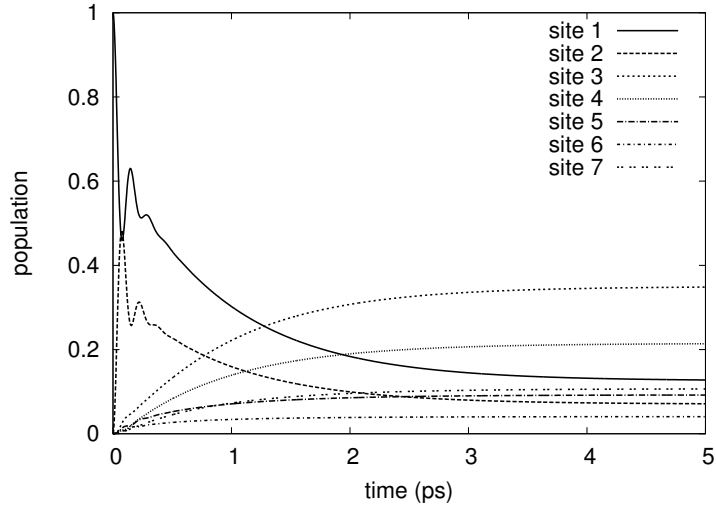}
\caption{\label{fig:popdyn}
Population dynamics in the FMO complex for an initial population present at site $1$ for the spectral density $J_{DL}$ with 
$\lambda=35$~cm$^{-1}$, $\gamma^{-1}=50$~fs at $T=277$~K. 
The thermalization process takes several picoseconds.
}
\end{figure}
This layout suggests that one important transport-mechanism is the relaxation toward the thermal equilibrium state with its higher occupation of energetically low-lying sites.
The thermalization process is visible in the population dynamics, Fig.~\ref{fig:popdyn}, where we initialize the system density matrix at time $t=0$ to a fully populated site 1 $\rho(t=0)=1\;|1\rangle\langle 1|$ and the bath is in a thermal equilibrium state.
Once the system thermalizes the final system density matrix at $t\rightarrow\infty$ is approximately given by
\begin{equation}\label{eq:rhothermal}
\rho_{\rm th}\approx\frac{\rme^{-H_{\rm ex}/k_B T}}{\tr\rme^{-H_{\rm ex}/k_B T}},
\end{equation}
where we ignore the entanglement of the system with the bath.
For the seven-site FMO complex and a Drude-Lorentz spectral density ($\lambda=35$~cm$^{-1}$, $\gamma^{-1}=50$~fs) Fig.~\ref{fig:popdyn} shows the approach to the thermal state after several picoseconds.
To study the transfer-process in detail, we add the loss-channel and the reaction center to the system Hamiltonian \cite{Kreisbeck2011a}.
One important figure of merit is the transfer time for an initial population at either site one or six to the reaction center given by \cite{Kreisbeck2011a}
\begin{equation}\label{eq:transfertime}
\langle t \rangle=\int_0^{\infty} \rmd t'\ t'\, \big(\frac{\rmd}{\rmd t} \langle RC | \rho(t) | RC \rangle \big)_{t=t'}
\end{equation}
To see how the transfer times depends on the strength of the coupling of electronic and vibrational parts of the system, we scan through increasing values of the reorganization energy $\lambda$ of a Drude-Lorentz spectral density and record the transfer time.
Fig.~\ref{fig:fmolevels} depicts the transfer time for different levels of approximation to the time-evolution of the density matrix. 
The exact ME-HEOM calculation reveals that a shortest transfer time around $\lambda=50$~cm$^{-1}$ exists, while the Redfield approach does not find the optimal minimum value (the generalized Bloch-Redfield approach can recover the optimal value \cite{Wu2010a}).
\begin{figure}[t]
\includegraphics[width=0.45\textwidth]{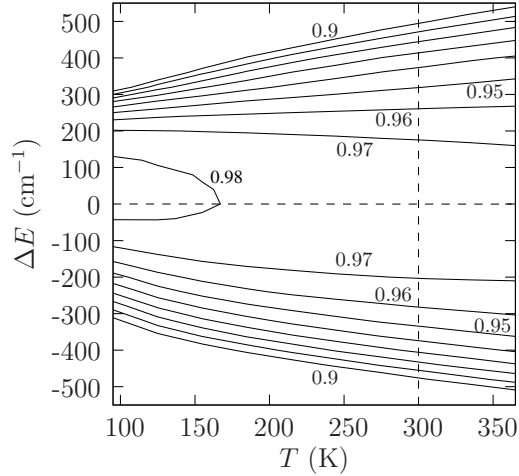}
\caption{\label{fig:FMOlevelshift}
Efficiency, eq.~(\ref{eq:eff}), of the transport through the FMO network from initial site 1 to the reaction center as function of temperature and shifting the energies of sites 3+4 (adapted from \cite{Kreisbeck2011a}). 
The dashed lines mark the zero-detuning case at temperature $T=300$~K. 
}
\end{figure}
Fig.~\ref{fig:FMOlevelshift} displays the efficiency of the transport, given by the long-time limit of the population of the reaction center (RC)
\begin{equation}\label{eq:eff}
\eta=\langle RC | \rho(t\rightarrow\infty) | RC \rangle
\end{equation}
as function of temperature and displacement of the site energies three and four. 
Detuning these two levels either further upwards or downwards from their unshifted positions diminishes the transport efficiency, since it is detuning sites 3 and 4 away from the rest of the system.
While a shift toward lower energy in principle enhances the thermal population of sites 3 and 4 the detuning suppresses the excitation flow into the sites and results in a decreased efficiency.

\section{FMO complex: comparison of theory and experiment}

To assess how closely the theoretical models match the experimental data requires to compute the experimentally measured quantities. 
In particular the time-resolved 2d echo-spectra provide stringent tests for theoretical results since they allow one to track the exciton dynamics in detail and thus to monitor the time-evolution of the energy-transfer from a specific initial condition.

The number of energy-levels which contribute to a two-dimensional spectrum scales as $D=(N_\text{sites}^2+N_\text{sites})/2+1$, which yields for a dimer $D=4$, for a trimer already $D=7$ and for a seven site model of the FMO complex $D=29$. 
The larger size of the system significantly affects the excitation flow. 
This implies that results obtained for dimer models regarding long-lasting oscillations and coherences cannot be readily extrapolated to the FMO complex.
Up to date, the hierarchical equations of motions method are the only method available for computing two-dimensional spectra of the FMO complex based on the Hamiltonian specified in eq.~(\ref{eq:Htot}), without resorting to the Redfield approximation \cite{Bruggemann2007}.

\begin{figure}
\includegraphics[width=0.6\textwidth]{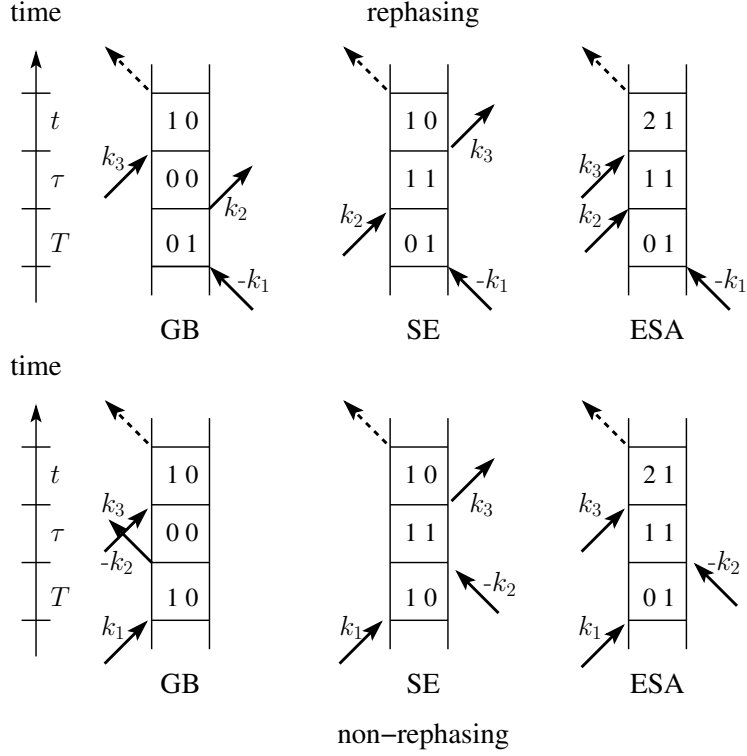}
\caption{\label{fig:liouvillepathways}
Representation of the light-matter interaction sequences (Liouville pathways) encountered in the third order response function $S^{(3)}(t,\tau,T)$. 
The upper panels show the rephasing pathways for ground-state bleaching (GB), stimulated emission (SE) and excited state absorption (ESA), the lower panels the corresponding non-rephasing ones.
The numbers denote the number of excitations present in the bra and ket of the system density matrix.
Time is proceeding upwards, the $k$ vectors denote the direction of the light pulses. 
Under phase-matching conditions the final signal pulse is emitted in direction $-k_1+k_2+k_3$ for the rephasing signal and into $+k_1-k_2+k_3$ for the non-rephasing case.
}
\end{figure}

\subsection{Computation of two-dimensional spectra}

Two-dimensional echo spectra are monitoring the third-order polarisation $P^{(3)}(t)$, which is given by the product of the commutator of the dipole operator $\hat{\mu}$ with the product of the electric field values ${\cal E}$ \cite{Mukamel1999a,Hamm2011a}
\begin{equation*}
P^{(3)}(t)=\int_0^\infty\rmd t_3\int_0^\infty\rmd t_2\int_0^\infty\rmd t_1
{\cal E}(t-t_3){\cal E}(t-t_3-t_2){\cal E}(t-t_3-t_2-t_1) S^{(3)}(t_3,t_2,t_1),
\end{equation*}
where we introduce the 3rd order response function 
\begin{equation}\label{eq:S3}
S^{(3)}(t_3,t_2,t_1) = \left(-\frac{\rmi}{\hbar} \right)^3 \tr\left(\mu\left(t_3+t_2+t_1\right) \left[\mu \left(t_2+t_1\right),\left[\mu\left(t_1\right),\left[  \mu (0),\rho_0\right]\right]\right] \right).
\end{equation}
The total incoming electric field is the superposition of three fields
\begin{equation}
{\cal E}(t)=\sum_{i=1}^3{\cal E}_i(t) \left( \rme^{-\rmi\omega_i t+\rmi k_i r}+\rme^{+\rmi\omega_i t-\rmi k_i r} \right).
\end{equation}
Assuming $\delta$ shaped pulse envelopes with ${\cal E}_1(t)={\cal E}_1\delta(t+T+\tau)$, ${\cal E}_2(t)={\cal E}_2\delta(t+\tau)$, and ${\cal E}_3(t)={\cal E}_3\delta(t)$ allows one to perform the $t_1, t_2, t_3$ integrations. 
If in addition a strict time-ordering of the three pulses is maintained and we only consider frequency combinations with are near the excitonic energy differences (rotating wave approximation), only two products of the electric fields remain
\begin{eqnarray}
\rme^{-\rmi [ (+k_1-k_2-k_3) r + (\tau+T) \omega_1 -\tau\omega_2]} S^{(3)}_{\rm RP}(t,\tau,T),\\
\rme^{-\rmi [ (-k_1+k_2-k_3) r - T \omega_1 +\tau(\omega_1-\omega_2)]} S^{(3)}_{\rm NR}(t,\tau,T).
\end{eqnarray}
The difference in $k$ vectors signifies different signal directions $k_{4}=\mp k_1\pm k_2 + k_3$ and makes a separate spatial detection possible (referred to as rephasing (RP) and non-rephasing (NR) respectively).
If all pulses have the same frequency $\omega=\omega_1=\omega_2$, which is often the case experimentally, the two terms simplify for $r=0$ to
\begin{equation}
\rme^{-\rmi T \omega} S_{\rm RP}^{(3)}(t,\tau,T),\quad \rme^{+\rmi T \omega} S^{(3)}_{\rm NR}(t,\tau,T).
\end{equation}
The 2d-echo spectra are determined by applying a Fourier transforms with respect to the first and last time-intervals.
We multiply with $\rme^{+\rmi t \Omega}$ to obtain the final expression for computing the amplitude of 2d echo-spectra
\begin{eqnarray}
I_{RP}^{(3)}(\Omega,\tau,\omega)=\int_0^\infty\rmd t\int_0^\infty\rmd T \; \rme^{+\rmi t \Omega-\rmi T \omega} S^{(3)}_{\rm RP}(t,\tau,T),\\
I_{NR}^{(3)}(\Omega,\tau,\omega)=\int_0^\infty\rmd t\int_0^\infty\rmd T \; \rme^{+\rmi t \Omega+\rmi T \omega} S^{(3)}_{\rm NR}(t,\tau,T).
\end{eqnarray}
\begin{figure}
\includegraphics[width=0.495\textwidth]{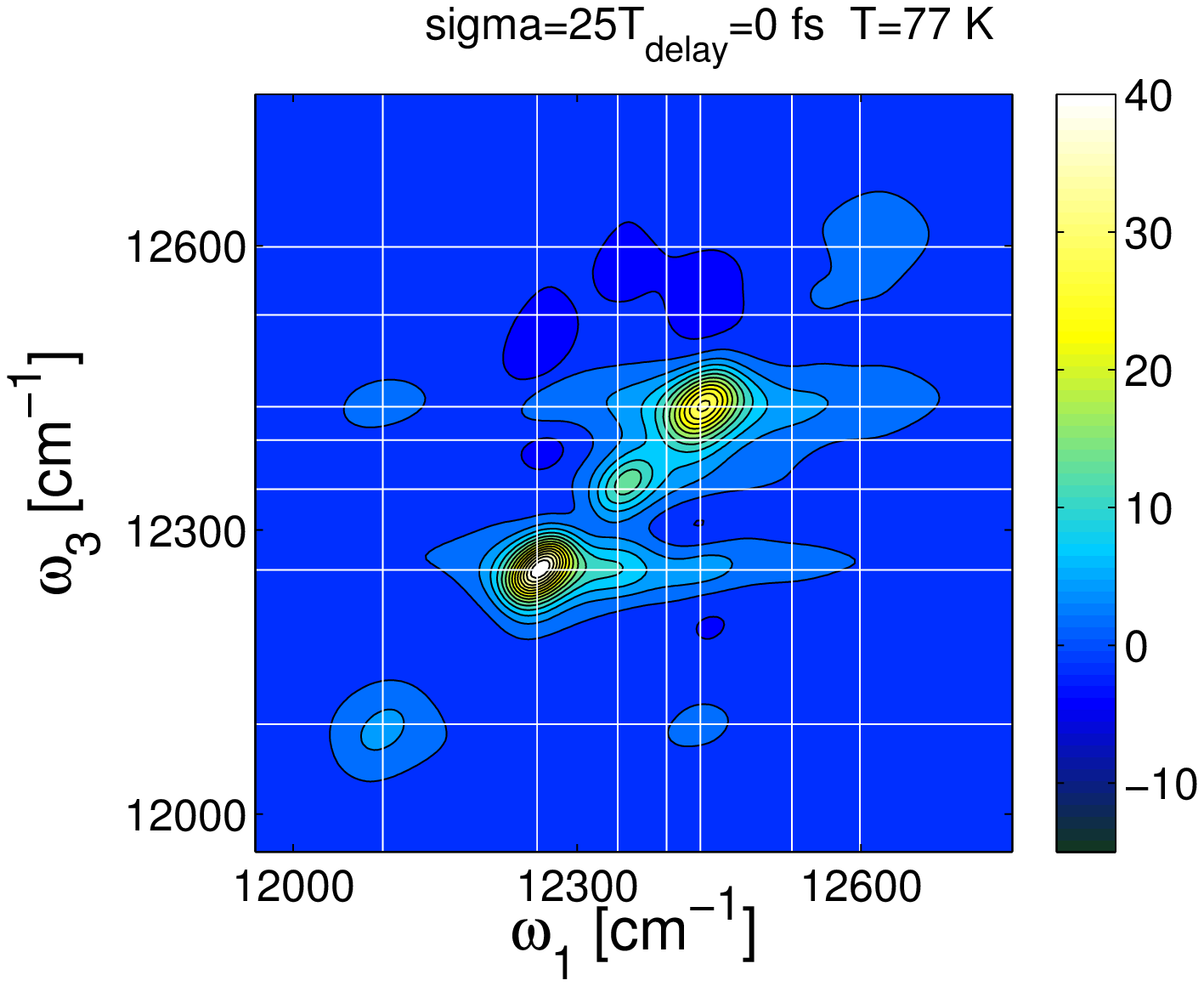}
\includegraphics[width=0.495\textwidth]{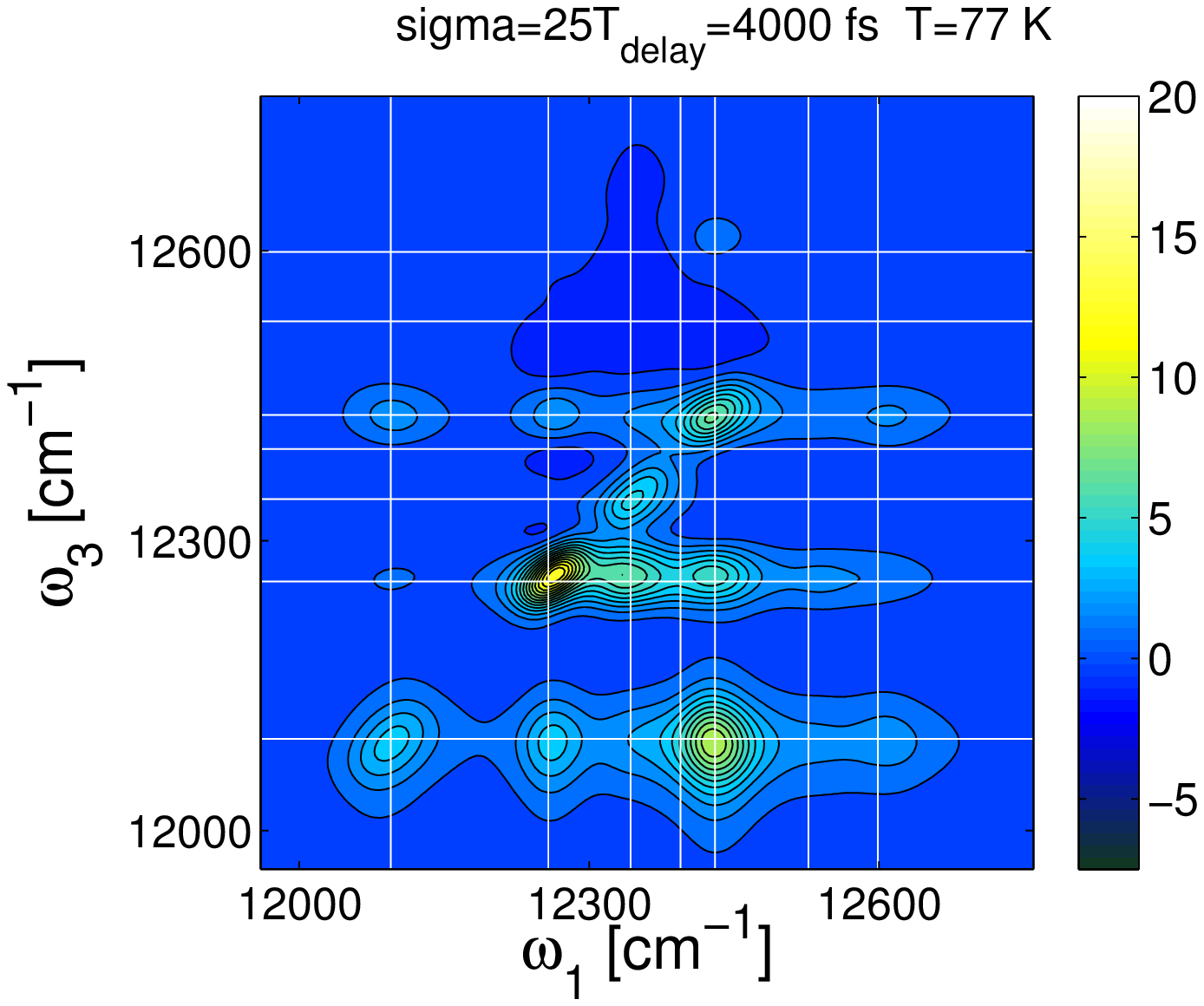}
\caption{\label{fig:2dDL} (color online)
Calculated two-dimensional echo-spectra for the FMO complex at $T=77$~K for delay time $T_{\rm delay}=0$~ps (left panel) and $T_{\rm delay}=4$~ps (right panel) for a Drude-Lorentz spectral density ($\lambda=35$~cm$^{-1}$, $\gamma^{-1}=50$~fs) under consideration of all six Liouville pathways and including static disorder and rotational averaging.
Note the movement of the initially diagonal amplitude to the lower frequency due to the dissipative coupling to the vibrational modes
(adapted from \cite{Hein2012a}).
}
\end{figure}
Thus the calculation is reduced to evaluating the 3rd order linear response function $S^{(3)}(t,\tau,T)$ given in terms of the dipole operator applied to the density matrix of the system at different times.
The dipole operator is readily constructed in site basis, where the direction of the transition dipole moment at site $k$ is denoted by $\vec{d}_k$. For a laser polarization along direction $\vec{l}$ this results in a dipole moment $\mu_k=\vec{d}_k\cdot \vec{l}$.
The dipole operator $\hat{\mu}$ for creating an excitation $\hat{\mu}^{+}$ or de-excitation $\hat{\mu}^{-}$ of the system from a ground state $|0\rangle$ added as site $0$ to the system then reads
\begin{equation}
\hat{\mu}^-=\left(
\begin{array}{ccccc}
0 & \mu_1  & \mu_2 & \cdots \\
0 & 0  & 0 & \cdots \\
0 & 0  & 0 & \cdots \\
\vdots & \vdots  & \vdots & \ddots
\end{array}
\right),\quad
\hat{\mu}^+={\left(\hat{\mu}^-\right)}^T,\quad
\hat{\mu}=\hat{\mu}^{+}+\hat{\mu}^{-}.
\end{equation}
Inserting this equation into the commutator and final product of the dipole operators eq.~(\ref{eq:S3}) and sorting the terms according to the non-rephasing and rephasing wave-vector requirement reduces the resulting sum to two sets of three terms (ground-state bleaching [GB], stimulated emission [SE], and excited-state absorption [ESA]), represented in terms of Liouville pathways in Fig.~\ref{fig:liouvillepathways}. In the diagrams, arrow pointing to the right (left) stands for an electric field amplitude of $\rme^{-\rmi \omega t+ \rmi k r}$ ($\rme^{+\rmi \omega t- \rmi k r}$), and the vertical lines represent the bra and ket of the density matrix. An arrowhead ending in the left (=ket) vertical line signifies an excitation with $\hat{\mu}^+$, while one pointing away stands for emitted light $\hat{\mu}^-$ (for the right (=bra) line, the reverse holds).
This representation allows one to apply the GPU-HEOM method for the propagation of the density matrix to the calculation of 2d spectra.
For instance the stimulated emission rephasing pathway is given by
\begin{equation}
S^{(3)}_{\rm RP,SE}(t,\tau,T)=\tr \left( \mu^-(\tau+T+t)\mu^+(\tau)\rho(0)\mu^-(0)\mu^+(\tau+T)\right).
\end{equation} 
The ESA pathways results from multiple excitation events, which let two excitons simultaneously propagate in the system. 
This requires to extend the single-exciton Hamiltonian and the dipole operators to the two-exciton manifold, which are in first approximation given by the addition of the corresponding single-exciton energies and dipoles, see \cite{Cho2006a,Hein2012a}.
For the seven-site FMO complex the addition of the ground-state and the two-exciton manifold requires to construct a $29\times 29$-density matrix.
The transition dipole moments $\vec{d}_k$ in the FMO complex are approximately directed along two of the four nitrogen atoms (denoted by $N_B-N_D$) in the Bacteriochlorophylls, with positions taken from the structure deposited in the protein data bank \cite{Tronrud2009a}.
Opposed to the regular and periodic structure found in the crystallized FMO complex, the experimentally probed sample contains an ensemble of many complexes, where each one is affected by a slowly changing protein environment (the protein scaffold is not shown in Fig.~\ref{fig:greensulfur}, it can be considered as quasi-stationary on the time-scale of the exciton propagation). 
This gives rise to set of different site-energies found in a statistical ensemble of the FMO complex and is referred to as static disorder.
In addition to the static disorder, the proteins are randomly oriented with respect to the laser fields, which requires to evaluate the rotational average as well.
To this end we perform ``10-shot'' simulations, which approximate the full sphere by a icosahedron \cite{Hein2012a}.
Both averages prolong computation times by roughly a factor $10^3$, since disorder ensembles encompass hundreds of computations, further multiplied by the rotational sampling. 
The last missing part for the evaluation of the 2d spectra is the specification of the spectral density $J(\omega)$ of the vibrational baths coupled to each site.

\subsection{Comparison with experimental data}
\begin{figure}
\includegraphics[width=0.55\textwidth]{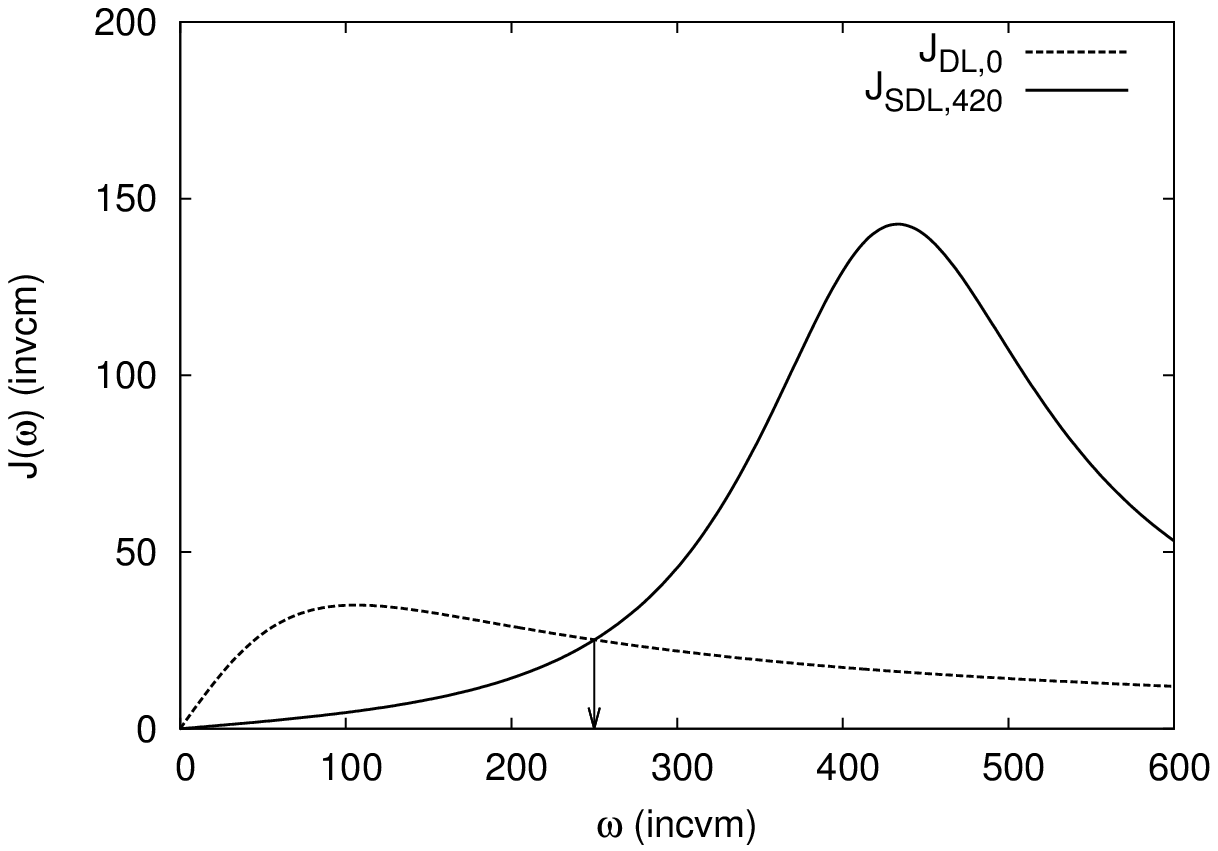}
\includegraphics[width=0.44\textwidth]{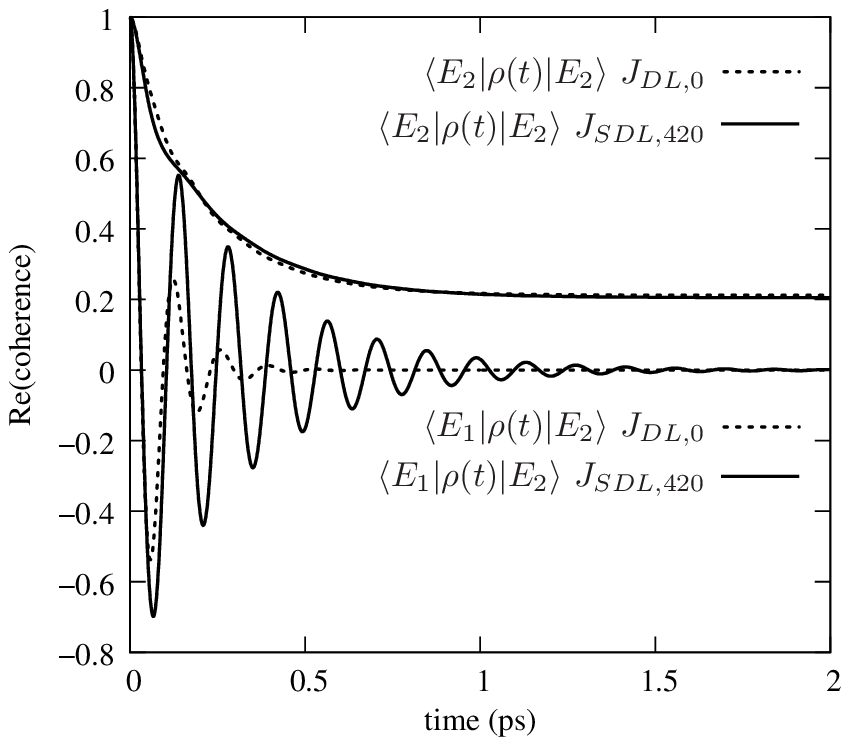}
\caption{\label{fig:relcoh}
Left panel:
Spectral density $J_{\rm DL,0}$ (unshifted Drude-Lorentz form, $\lambda=35$~cm$^{-1}$ and $\nu^{-1}=50$~fs) and $J_{\rm SDL,420}$ 
(shifted Drude-Lorentz peak, $\Omega=420$~cm$^{-1}$, $\lambda=35$~cm$^{-1}$ and $\nu^{-1}=50$~fs).
The arrow indicates the difference of eigenenergies of $H_{\rm ex}$ eq.~(\ref{eq:dimer}), where by construction both spectral densities have the same value.
Right panel:
Relaxation of the diagonal element $\langle E_2| \rho(t) |E_2 \rangle$ to the thermal state (upper non-oscillatory graphs) and damped oscillations of the off-diagonal coherence ${\rm Re} (\langle E_1| \rho(t) |E_2 \rangle)$ at $T=277$~K.
While both spectral densities give very similar relaxation rates, the off-diagonal coherence is much prolonged for $J_{\rm SDL,420}$ due to its small slope toward $\omega\rightarrow 0$.
}
\end{figure}

The two-dimensional spectra allow one to follow the exciton flow through the network by varying the second delay time $T_{\rm delay}=\tau$ between the pulses, while the Fourier transform over the first and last time-intervals are providing a two-dimensional frequency grid.
The horizontal frequency axis ($\omega_1=\omega$) corresponds to the excitation caused the first pulse, while the ($\omega_3=\Omega$) axis displays the frequency of the emitted light after delay time $T_{\rm delay}$. 
On the $(\omega_1,\omega_3)$-grid amplitude-peaks are expected at the location of exciton eigenenergies along the diagonal and cross-peaks at the intersection of eigenstates $E_i$ and $E_j$ 
If during the delay time energy is dissipated from the electronic part of the system to the vibrational bath, the electronic amplitude will move from the diagonal axis (on which absorbed and emitted frequencies coincide) to the lower diagonal space.
Such a relaxation as function of delay time has been observed by Brixner et al.\ \cite{Brixner2005a} and provides an important experimental verification of energy-funneling in the FMO complex.
On the theoretical side, this process is captured in 2d-spectra, shown in Fig.~\ref{fig:2dDL} for a simple $J_{DL}$ spectral density \cite{Hein2012a}.
The long-time limit of the time-evolution of the density matrix approaches the thermal state (\ref{eq:rhothermal}), up to the small system-bath entanglement.
In newer experiments, on top of the relaxation process, long-lasting oscillations are seen \cite{Engel2007a}, which are matching the excitonic energy differences \cite{Caram2011a} and are not affected by changing the vibrational density for example by replacing hydrodgen atoms with deuterium \cite{Hayes2011a}.
In a better controllable dimer system \cite{Hayes2013a}, the known electronic frequencies of the constituents are reflected in a periodic oscillations of the 2d echo-spectra of the dimer with period matching the energy-differences.
\begin{figure}
\includegraphics[width=0.6\textwidth]{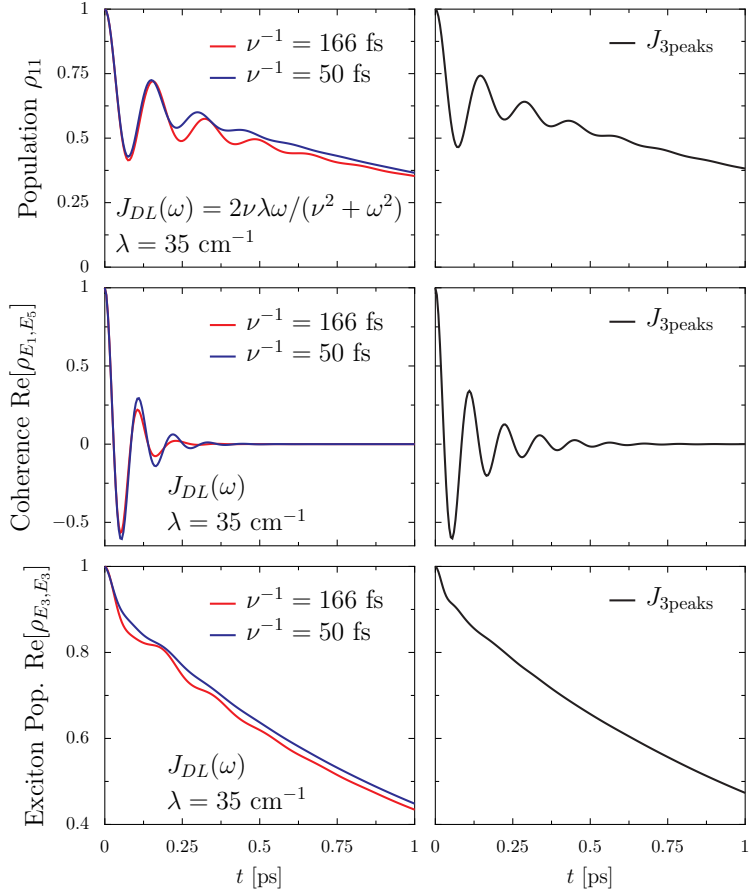}
\caption{\label{fig:cohdl}
(color online)
Coherence life-times and population beatings for different models of the spectral density. Both densities $J_{\rm 3peaks}, J_{\rm DL}$ display population beatings lasting up to $1$~ps.
However only $J_{\rm 3peaks}$ entertains long-lasting coherence in the off-diagonal element $\rho_{E_1,E_3}$, while the long-time bath correlations in $J_{\rm DL}$ lead to oscillations of the diagonal eigenstate population $\rho_{E_3,E_3}$.
Parameters: $T=150$~K, $J_{\rm DL}$ with $\lambda=35$~cm$^{-1}$, the value of $\gamma^{-1}$ is given in the panels.
Adapted from the supplementary information of Ref.~\cite{Kreisbeck2012a}.
}
\end{figure}

In experiments oscillations are visible for up to $2$~ps at $T=77$~K and still discernible at $T=277$~K \cite{Panitchayangkoon2010a}.
An initial hypothesis for the existence of long-lasting oscillations was the presence of long-lasting memory effects of the bath \cite{Ishizaki2009a}.
The effect of a slow decay of the bath correlation function has been seen in beatings of the computed population dynamics of the FMO complex for models with a single-peak Drude-Lorentz spectral density $J_{DL}(\omega)$.
Later computations of 2d-spectra showed that memory effects are causing oscillations, however at diagonal peak amplitudes and not at the experimentally determined cross-peaks \cite{Kreisbeck2012a}.
The failure of the long-memory model is related to the choice of $J_{DL}(\omega)$ as spectral density.
As instructive example, we consider the following dimer
\begin{equation}\label{eq:dimer}
H_{\rm ex}=\left(
\begin{array}{cc}
-75 & 100 \\
100 & 75 
\end{array}
\right)\;{\rm cm}^{-1}
\end{equation}
and study the time evolution of the system density-matrix initialized to (i) the highest eigenstate $\rho(t=0)=|E_2\rangle\langle E_2|$, and (ii) to a superposition  $\rho(t=0)=|E_1\rangle\langle E_2|$.
The diagonal element approaches the thermal state exponentially with the relaxation time $t_{\rm relax}$
\begin{equation}
\frac{\langle E_1 | \rho(t) | E_1 \rangle - \langle E_1 | \rho(\infty) | E_1 \rangle}{\langle E_1 | \rho(0) | E_1 \rangle - \langle E_1 | \rho(\infty) | E_1 \rangle} = \rme^{-t/t_{\rm relax}},
\end{equation}
while the off-diagonal coherences oscillate and decay with a decoherence-time $t_{\rm decoh}$
\begin{equation}
\langle E_1 | \rho(t) | E_2 \rangle = \rme^{-\rmi (E_2-E_1) t/\hbar}\rme^{-t/t_{\rm decoh}},
\end{equation}
\begin{figure}[t]
\includegraphics[width=0.9\textwidth]{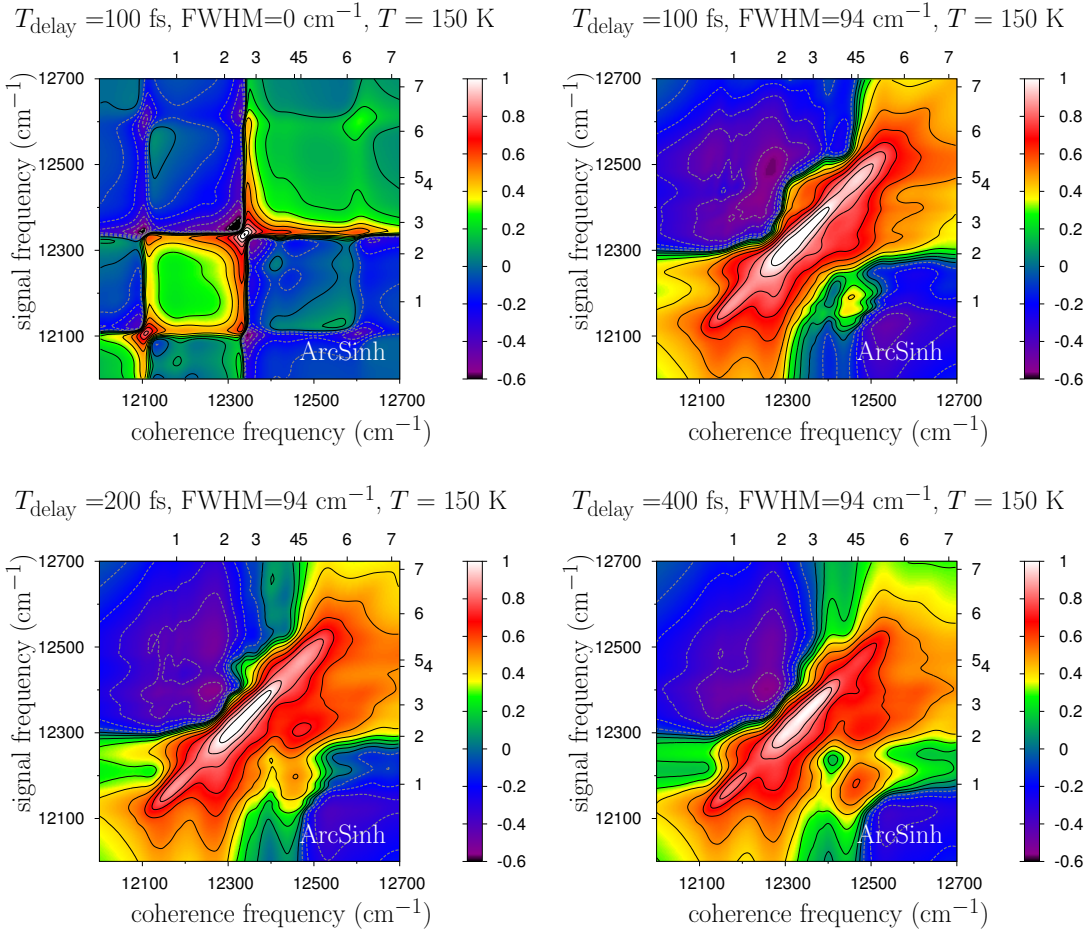}
\caption{\label{fig:2dfmoj3}
(color online) Computed 2d-echo spectra of the FMO complex at $T=150$~K without disorder (upper left panel) and for various delay times with static disorder added at the diagonal sites of the exciton Hamiltonian with a Gaussian distribution ${\rm fwhm}=94$~cm$^{-1}$ (all other panels).
The amplitude is scaled with an ${\rm arcsinh}$ function to facilitate comparison with experimental data.
A downward movement of amplitude from the diagonal axis towards the lower right frequency region is clearly visible, see for instance the location of the $(1,5)$ crosspeak.
}
\end{figure}
\begin{figure}
\includegraphics[width=0.9\textwidth]{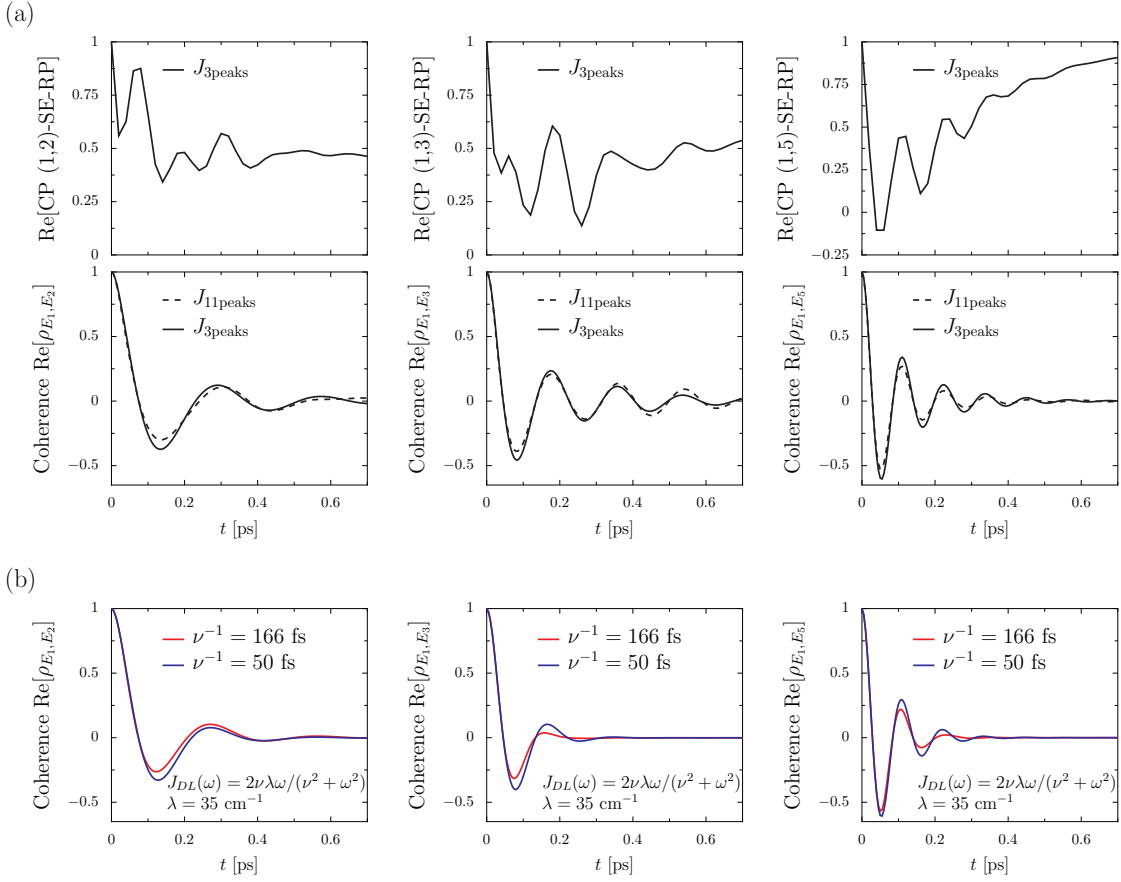}
\caption{\label{fig:2doscillations}
(a) upper panels: time-dependent peak amplitude of the rephasing stimulated-emission pathway at different cross-peak locations in the 2d-echo spectra calculated for the spectral density $J_{\rm 3 peaks}$ at temperature $T=150$~K, no disorder average.
Lower panels: time-evolution of the corresponding coherences, demonstrating good agreement with the decoherence rates found in stimulated-emission 2d-spectra.
(b) comparison with a single-peak Drude-Lorentz spectral density, which damps coherences on a much faster time-scale.
Adapted from the supplementary information of Ref.~\cite{Kreisbeck2012a}.
}
\end{figure}

Both decay rates are determined by the spectral-density, in particular the slope of $J(\omega)$ toward zero frequency and the value of $J(\omega)$ at the eigenenergy difference $E_2-E_1=250$~cm$^{-1}$.
The spectral densities $J_{DL,0}$ is identical to the unshifted Drude-Lorentz one in eq.~(\ref{eq:JDL}), while $J_{SDL,420}$ is given by eq.~(\ref{eq:SDL}) with the shift $\Omega=420$~cm$^{-1}$. 
Both spectral densities have the same value at the eigenenergy difference $250$~cm$^{-1}$, marked by the arrow in the left panel of Fig.~\ref{fig:relcoh}, and the same reorganization energy $\lambda=35$~cm$^{-1}$ and bath-correlation time $\nu^{-1}=50$~fs.
For a weak system-bath coupling, analytical estimates based on the spin Boson model are possible \cite{Weiss2008a} and show that the value of the spectral density at the eigenenergy difference $J(E_2-E_1)$ is responsible for the relaxation toward the thermal state of the diagonal element of the density matrix. 
Since we have constructed a case where $J_{\rm DL,0}(E_2-E_1)=J_{\rm SDL,420}(E_2-E_1)$, we find close agreement of the relaxation times $t_{\rm relax}$ shown in the right panel of Fig.~\ref{fig:relcoh}, upper (non-oscillatory) lines.
The decoherence rate is partly tied to the relaxation rate, but in addition strongly depends on the slope of the spectral density at zero frequency $J'(\omega=0)$ which sets the pure-dephasing rate.
The slope is largely reduced for $J_{\rm SDL,420}$ due to the shift $\Omega=420$~cm$^{-1}$ and leads to much longer-lived coherences (lower oscillatory graphs in the right panel of Fig.~\ref{fig:relcoh}).
The dimer example demonstrates the important influence of the slope of the spectral density toward zero frequency.
A very similar analysis and conclusion holds for the seven-size FMO complex shown in Fig.~\ref{fig:cohdl} for the unshifted Drude-Lorentz spectral density and for $J_{\rm 3peaks}$, which  follows the smooth part of the experimentally measured spectral density $J_{\rm Wendling}$  (Fig.~\ref{fig:specdens}).
The lower panels in Fig.~\ref{fig:cohdl} depict the approach to the thermal equilibrium and show a similar relaxation rate for the three spectral densities considered. 
The prolonged off-diagonal coherence lifetime of $J_{\rm 3 peaks}$ compared to $J_{DL}$ (middle panels) is caused by the smaller slope of  $J_{\rm 3 peaks}$ for $\omega\rightarrow 0$. 
Therefore we identify the small slope condition of the spectral density as important prerequisite for long-lasting coherences.
For the efficiency of the transport it is also important that the system evolves fast toward the thermal state, which is the case for all spectral densities (Fig.~\ref{fig:cohdl}, lower panels).

Next, we turn to the analysis of calculated 2d-echo spectra for the spectral densities $J_{\rm 3 peaks}$, see Fig.~\ref{fig:2dfmoj3}.
The panels depict the summed contributions of the rephasing and non-rephasing parts including ground-state bleaching and stimulated emission pathways at various delay-times, but not excited state absorption.
The inclusion of static disorder changes the 2d-spectrum considerably compared to the non-disorder case (upper left panel) and brings the theoretical result closer to the experimental data \cite{Panitchayangkoon2010a}, Fig.~1.
The theoretical calculation displays narrower peaks and differs from the results obtained for a single-peak Drude Lorentz spectral density, shown in Fig.~\ref{fig:2dDL}.
In Fig.~\ref{fig:2dfmoj3} the markers at the right and upper frequency axis denote the location of the exciton eigenenergies and allow to record the amplitude at specific cross-peaks as function of delay time \cite{Kreisbeck2012a}.
In contrast to the Drude-Lorentz case, the cross-peaks do undergo longer-lasting coherent oscillations. 

\begin{figure}
\includegraphics[width=0.8\textwidth]{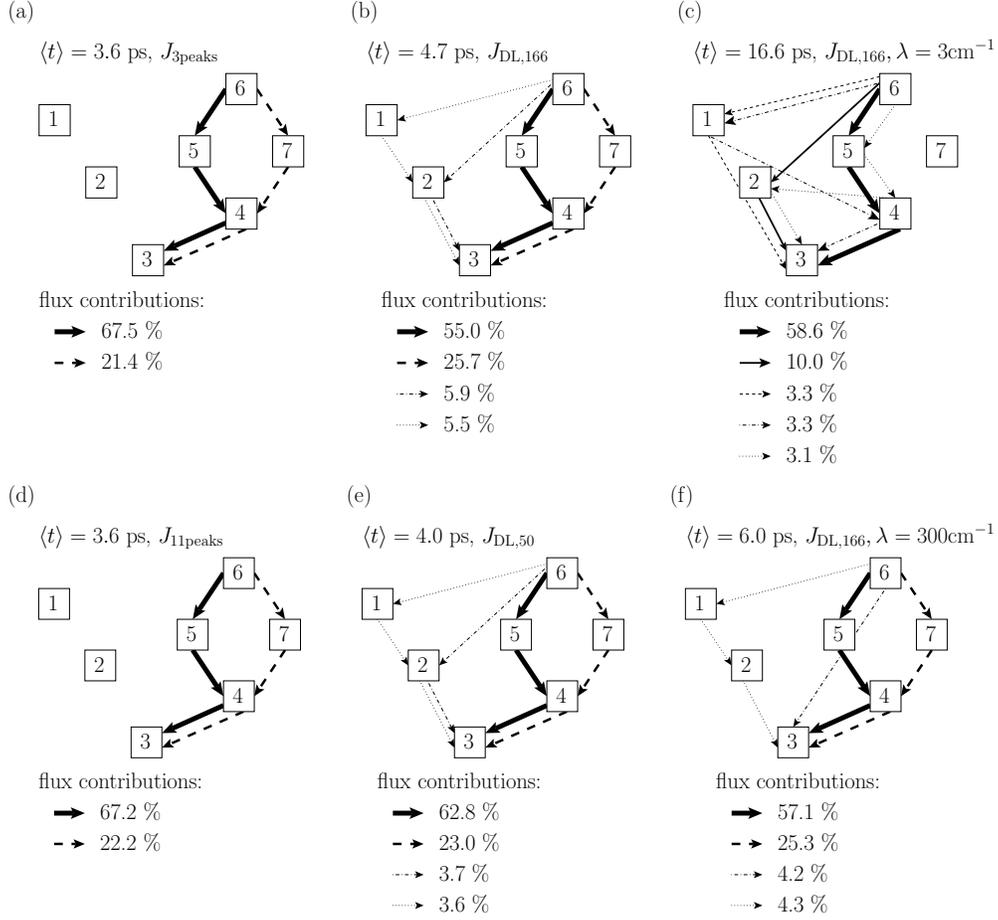}
\caption{\label{fig:fluxanalysis}
Flux analysis and dominant pathways of energy flow from initial population at site~6 to the reaction center linked to site~3 for different spectral densities (a) $J_{\rm 3peaks}$, (d) $J_{\rm 11peaks}$, and (b) $J_{\rm DL}$ ($\lambda=35$~cm$^{-1}$, $\nu^{-1}=166$ fs),
(c) $J_{\rm DL}$ ($\lambda=3$~cm$^{-1}$, $\nu^{-1}=166$ fs), (e) $J_{\rm DL}$ ($\lambda=35$~cm$^{-1}$, $\nu^{-1}=50$ fs), (f) $J_{\rm DL}$ ($\lambda=300$~cm$^{-1}$, $\nu^{-1}=166$ fs).
Shown are all pathways which contribute with more than 3\% to the total energy flux at temperature $T=300$~K and with trapping rate $\Gamma_{3\rightarrow RC}^{-1}=1$~ps. 
The spectral densities $J_{\{3,11\}\rm peaks}$ with small initial slope (a+d) both funnel the flux most direct to the reaction center.  
}
\end{figure}

In the theoretically calculated stimulated-emission [SE] pathways of 2d echo-spectra, the relaxation of diagonal peak amplitudes toward lower-lying eigenstates and periodic oscillations at off-diagonal peaks are visible on time-scales matching the simpler analysis of the coherences (Fig.~\ref{fig:2doscillations}).
Upon inclusion of all simultaneously observed pathways by adding exited state absorption [ESA] and ground-state bleaching [GB], the picture becomes more complicated, since the ESA pathways oscillates with a phase-shift by $\pi$ relative to the SE contribution, leading to a reduction of the oscillatory component.
In addition the GB pathway for a structured spectral density with narrow peaks in the vibrational spectral density contributes an oscillatory component with vibrational rather than electronic frequencies to the 2d-echo spectra \cite{Tiwari2012a,Kreisbeck2013b}. 
The ESA pathway describes the simultaneous propagation of two excitons in the system, while the ground-state bleach case induces no population within the system during the delay time $\tau$ (Fig.~\ref{fig:liouvillepathways}).
Both pathways are \textit{not relevant for the transport} of single excitons from the antenna to the reaction center, and thus the direct mapping of decoherence time-scales found in 2d-spectra to transport requires a careful assessment.

\subsection{Network flow}

The transfer time, eq~(\ref{eq:transfertime}), provides a global characterization of the overall transport process, but does not reveal in detail which routes through the network are most used. 
Guiding experimentalists in the fabrication process of new devices requires to unveil design principles and strategies for optimizing parts of the network. 
To this end, we discuss how the shape of spectral density selects specific pathways for the energy flow toward the trapping 
target site.
In the FMO complex the two possible initial sites (site 1 and site 6) are coupled to two dominant transport pathways (see Fig.~\ref{fig:fmolevels}). 
The first pathway involves site~1 and site~2 to transfer energy to site~3 which is coupled to the reaction center. 
The second pathway follows sites~6, 5, 7, and 4 before reaching site~3 and then the reaction center.
Formulated in terms of an optimization problem, we need to find a Hamiltonian and couplings to the environment which make the transfer most efficient. 
In particular two criteria need to be fulfilled:
(i) the energy needs to find the shortest way to the reaction center. 
Any random detours through the network need to be avoided since they incur losses. 
(ii) the two pathways should be of similar efficiency to account for the redundancy. 
Both criteria are met by aligning the energy levels from the initial sites to the reaction center in the form of an energy funnel.
The downhill transfer is then assisted by the dissipative environment and constitutes the most dominant effect for efficient energy transfer in the FMO complex. 
Since the coupling to the environment plays such an important role we expect that the shape of the spectral density, 
which parametrizes this coupling, has a significant impact on optimizing transport properties.
\begin{figure}
\includegraphics[width=0.6\textwidth]{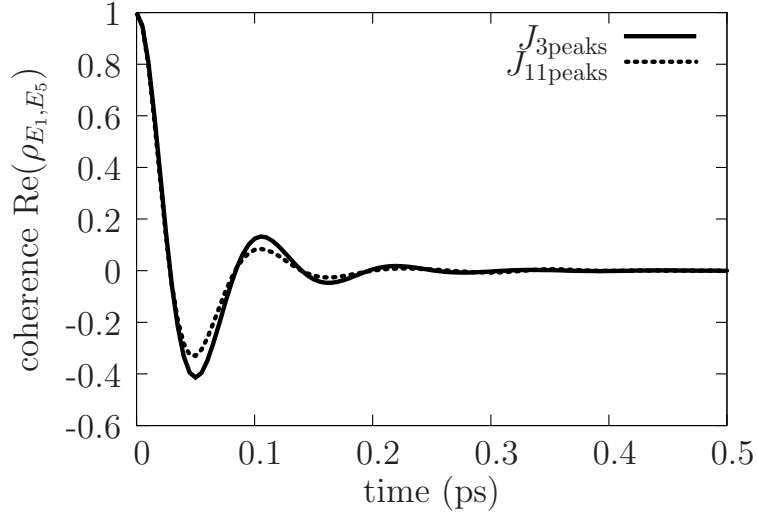}
\caption{\label{fig:j3j11}
Coherences of the exciton eigensates $E_1$ and $E_5$ of the FMO complex for the differently structured spectral densities $J_{\rm 3 peaks}$ vs $J_{\rm 11 peaks}$ at $T=300$~K, showing very similar decay rates of the electronic coherence.
}
\end{figure}

To show this, we analyze the flux through the FMO complex for various spectral densities.
Following \cite{Wu2012a}, the quantum mechanical flux from site~$m$ to site~$n$ is defined as
\begin{equation}
 F_{nm}=2{\rm Im}\{J_{nm}\tau_{mn}\}
\end{equation}
with
\begin{equation}
 \tau_{mn}=\int_0^\infty \mbox{d}t\ \langle m|\rho(t)| n\rangle.
\end{equation}
The flux is normalized according to 
\begin{equation}
 q_{mn}=\frac{F_{mn}}{\sum_{m', F_{m,m'}>0}F_{mm'}}
\end{equation}
to determine the relative contribution of the individual pathways to the total transport through the network. 

Figure~\ref{fig:fluxanalysis} shows the dominant pathways of energy flow in the FMO complex at $T=300$~K for $J_{\rm 3peaks}$, $J_{\rm 11peaks}$, and $J_{\rm  DL}$ for different reorganization energy $\lambda$ and bath correlation time $\nu$. 
The initial population is located at site~6 and we neglect the small amount of energy losses due to recombination.
The trapping rate is set to $\Gamma_{3\rightarrow RC}=1$~ps for the coupling of the reaction center to site~3 of the FMO complex. 

The spectral densities $J_{\rm 3peaks}$ and $J_{\rm 11peaks}$ show a very similar flux distribution, which directs the flow efficiently along the direct routes.
For $J_{\{3,11\} \rm  peaks}$ $67\%$ of the total flux proceeds along sites~6,5 and 4 and 3 to the reaction center, while $22\%$ reaches the target through sites~6, 7, 4 and 3. 
For $J_{\rm DL}$ the situation is different. 
Although the main contributions go along the shortest pathways, significant parts are diverted to sites~1 and 2 before the energy reaches site~3 and the reaction center. 
This less-efficient distribution holds for a large range of coupling strengths. 
For almost vanishing dissipation (c), the excitation explores the complete network rather than being directed to the reaction center.

The striking similarity of the flow patterns of $J_{\{3,11\} \rm  peaks}$ shows that the energy transfer is not affected by added peaks at the excitonic eigenenergy differences.
This finding is in contrast to speculations in the literature which single out those vibronic modes as fundamental design principle.

\section{Summary of design lessons from nature: fine-tuning is futile}

The experimentally derived spectral density $J_{\rm Wendling}$ in Fig.~\ref{fig:specdens} shows additional narrow-peaks, which are mostly contained in the parametrization $J_{\rm 11 peaks}$. 
To investigate wether these peaks play an important role for long-lasting oscillations requires to analyze the exciton-coherence for $J_{\rm 11 peaks}$ and to compare it to the less peaked spectral density $J_{\rm 3 peaks}$.
The result in Fig.~\ref{fig:j3j11} shows that the added narrow peaks of $J_{\rm 11 peaks}$ have only a minor impact ($\ll 0.01$) on the amplitude of the coherences.
In 2d-spectra this small influence can be magnified due to the different sensitivity of the pathways on the vibrational bath. 
For instance, ground-state bleach does not reflect any coherences in the system, but is sensitive to the spectral bath modes.
In experiments, the different pathways are measured simultaneously, while in theoretical calculation a separate analysis is possible and reveals the different time-scales and frequencies involved in vibrational beatings compared to electronic coherences \cite{Kreisbeck2013b}.
\begin{figure}
\includegraphics[width=0.53\textwidth]{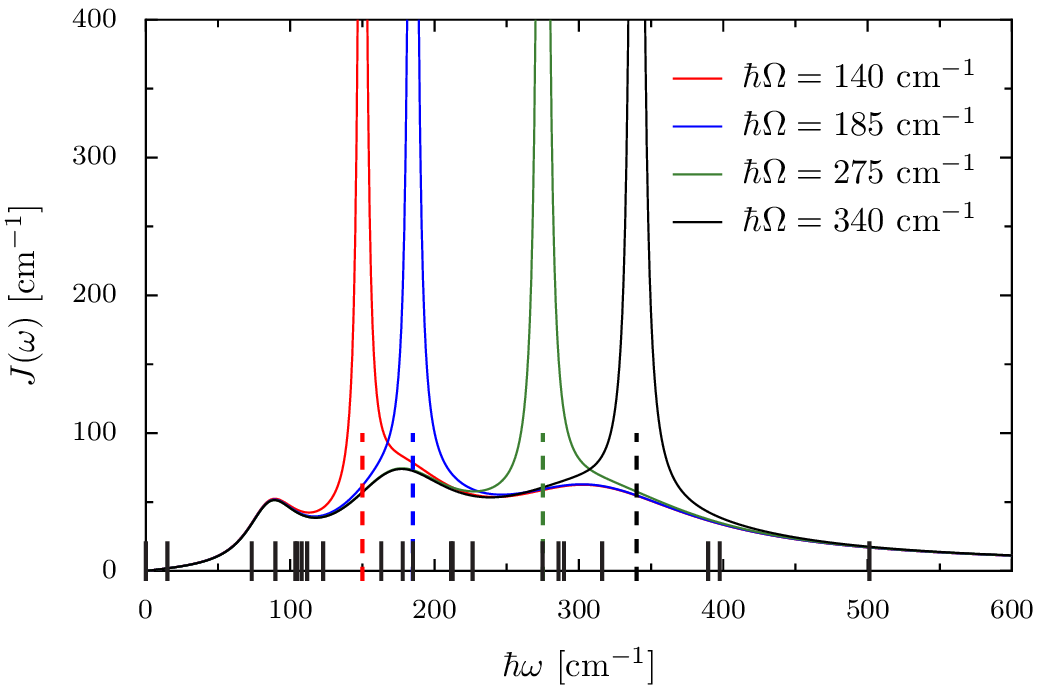}
\includegraphics[width=0.46\textwidth]{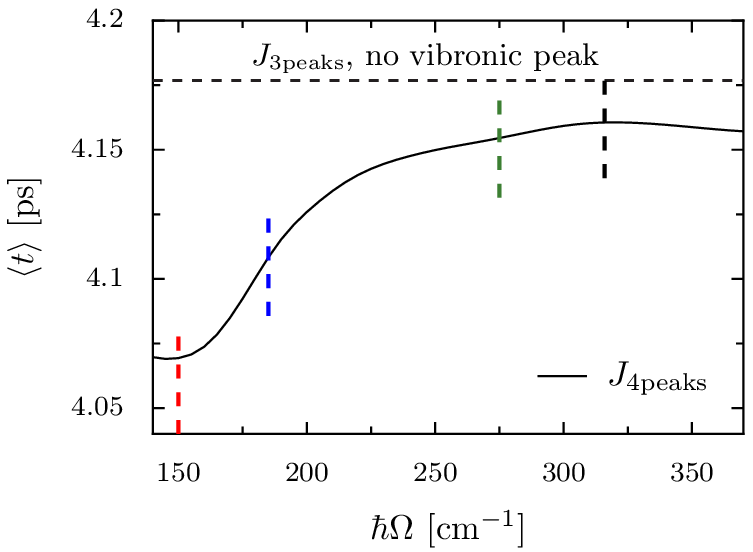}
\caption{\label{fig:peakscan}
(color online)
Left panel: spectral density $J_{\rm 3 peaks}$ with one added peak ($\lambda=15.6$~cm$^{-1}$ , $\nu^{-1}=2000$~fs) shifted to different frequencies which overlap with exciton eigenenergy differences of the FMO complex, indicated by markers on the horizontal axis.
Right panel: transfer time as function of added peak position, showing only small overall changes with no noticeable faster transfer time when the peak is tuned to eigenenergy differences.
}
\end{figure}
To investigate how the transport efficiency is affected by the presence of additional peaks on top of the background spectral density, we compare in Fig.~\ref{fig:peakscan} the transfer time through the FMO complex with an added vibrational peak shifted through resonances with the electronic eigenenergy differences. 
The fine-tuning of the additional peak with respect to electronic eigenenergy differences does not result in a faster transfer time.
The underlying design principle for efficient transport is already realized with $J_{\rm 3 peaks}$ in a robust fashion and does not rely on specifically added resonant peaks, which would require to match vibrational frequencies to specific disorder realization. 
This conclusion is also supported by the network flow analysis comparing $J_{\rm 3 peaks}$ vs $J_{\rm 11 peaks}$, which shows negligible effects of resonant vibrational modes on the flow pattern through the network.

\begin{theacknowledgments}
TK thanks the organizers of the Latin American School of Physics Marcos Moshinsky for the invitation to present this course at El Colegio Nacional, M\'exico D.F. and the hospitality during this stay.
We acknowledge helpful discussions with Al\'an Aspuru-Guzik, Gregory Engel, and Jan Ol\v{s}ina.
TK is supported by a Heisenberg fellowship of the DFG, grant KR 2889/5-1 and CK by DARPA grants N66001-10-1-4059 and N66001-10-1-4063.
\end{theacknowledgments}

\bibliographystyle{aipproc}   

\end{document}